\documentclass[12pt,A4]{article}
\usepackage{psfig}
\usepackage{epsfig}

\def\gee{ \, \lower 1mm\hbox{$\,{\buildrel > \over{\scriptstyle\scriptstyle\sim} }\displaystyle \,$}}
\def\lee{ \, \lower 1mm\hbox{$\,{\buildrel < \over{\scriptstyle\scriptstyle\sim} }\displaystyle \,$}}
\def\Oo {\displaystyle}
\def\varkappa {{\scriptstyle\partial}\! e}

\let\b=\baselineskip
\let\c=\centerline

\begin{document}

\renewcommand{\baselinestretch}{1.4}

\hfill{\sc Journal-ref: Astronomy Reports, 2003, v.~47, No.~5,
pp.~357--376}
 \bigskip \bigskip

\c{\bf\Large Minimum Velocity Dispersion in Stable Stellar Disks.}
 \medskip

\c{\bf\Large Numerical Simulations}
 \bigskip

\c{\it \Large A.V. Khoperskov~$^1$, A.V. Zasov~$^2$, N.V.
Tyurina~$^3$}
 \bigskip

\noindent $^1$
 Sternberg Astronomical Institute, Universitetskii pr.
13, Moscow, 119992 Russia;

\noindent Volgograd State University;

\noindent khopersk@sai.msu.ru
 \smallskip

\noindent $^2$
 Sternberg Astronomical Institute, Universitetskii
pr. 13, Moscow, 119992 Russia;

\noindent zasov@sai.msu.ru
 \smallskip

\noindent $^3$
 Sternberg Astronomical Institute, Universitetskii
pr. 13, Moscow, 119992 Russia;

\noindent  tiurina@sai.msu.ru

\begin{abstract}{\large
We used $N$-body dynamical simulations to analyze the conditions
for the gravitational stability of a three-dimensional stellar
disk in the gravitational field of two rigid spherical components
--- a bulge and a halo, whose central concentrations and relative
masses vary over wide ranges. The number of point masses $N$ in
the simulations varied from 40 to 500 thousands and the evolution
of the simulated models is followed over 10--20~rotation periods
of the outer edge of the disk. The initially unstable disks are
heated and, as a rule, reach a quasi-stationary equilibrium with a
steady-state radial-velocity dispersion $c_r$ over five to eight
periods of rotation. The radial behavior of the Toomre stability
parameter $Q_T (r)$ for the final state of the disk is estimated.
Numerical models are used to analyze the dependence of the
gravitational stability of the disk on the relative masses of the
spherical components, disk thickness, degree of differential
rotation, and initial state of the disk. Formal application of
existing analytical local criteria for marginal stability of the
disk can lead to errors in $c_r$ of more than a factor of 1.5. It is suggested that the approximate
constancy of $Q_T \simeq 1.2\div 1.5$ for $r\simeq (1\div2)\times
L$ (where $L$ is the radial scale of a disk surface density), valid
for a wide range of models, can be used to estimate upper limits
for the disc mass and density based on the observed
distributions of the rotational velocity of the gaseous component
and of the stellar velocity dispersion. }
\end{abstract}

\vfill\eject

\section{INTRODUCTION}

{ \b=1.1\b

Galactic disks, which consist mostly of old stars, can be
considered as collisionless systems in quasi-stationary
equilibrium with very slow evolution. This state is characterized
by certain radial dependences of the stellar velocity dispersions
$(c_r,c_\varphi,c_z)$ that ensure stability of the disk. Knowledge
of the stability conditions makes it possible to develop
self-consistent models for the disks of real galaxies for which
both the rotational velocities and stellar velocity dispersions
have been measured at various galactocentric distances.

The problem of determining the minimum stellar velocity dispersion
sufficient to ensure stability of the disk against arbitrary
perturbations is especially important if, as many authors have
suggested (see, e.g.,~[1--7]), real galaxies may be in a state of
threshold (marginal) stability. This approach enables the local
density or integrated mass of the disk to be estimated from the
observed velocity dispersion. In the general case, the old stellar
population of a galactic disk can have an excess velocity
dispersion in the presence of other factors that heat the disk,
which are not directly related to the gravitational instability.
However, even in this case, the conditions for marginal stability
provide valuable information by yielding an upper limit for the
mass of the disk that enables it to be stable.

Fridman and Polyachenko~[8,~9] carried out a detailed theoretical
analysis of the stability of thin rotating disks against
different kinds of perturbations (including bending modes).
Together with certain advantages over numerical simulations (the
mathematical rigorousness of the solutions in the framework of
the problem formulated), the analytical approach to the dynamics
of perturbations in a gravitating disk and the conditions for

stability has the drawback that it can be implemented only for
very simple models and can yield only coarse estimates for the
parameters of the disk component when applied to real objects. Let
us consider these simplifications in more detail.

First and foremost, the simple analytical models of collisionless
disks that are used in stability analyses usually assume that the
disk thickness is small; they actually consider an infinitely thin
layer. Despite the smallness of the ratios $h_z/r$ and $h_z/L$
over most of the stellar disk (here, $r$ is the radial coordinate,
$h_z$ is the vertical scale height, and $L$ is the radial scale
length of the disk), this condition may not be sufficient to
justify neglect of vertical motions (see Gor'kavyi and
Fridman~[10, Appendix II\footnote{Written in coauthorship with
O.V.~Khoruzhi{\u\i}.}] for a detailed discussion of this issue; in
the general case, the dynamics equations for astrophysical disks
cannot be adequately treated in a two-dimensional formulation).

Analytical studies of the dynamics of small perturbations in a
stellar disk usually assume that the disk parameters have small
radial gradients (see, e.g.,~[1, 11, 12]). Attempts to allow for
gradients have thus far been made only in terms of WKB
approximations (see, e.g., [2, 13--15]). This means that the
characteristic perturbation wavelength is short compared to the
local scale lengths for variations of the circular velocity
$V(r)$, stellar radial-velocity dispersion $c_r(r)$, and surface

density $\sigma(r)$. In many cases, these conditions are barely
satisfied or even not satisfied. Taking into account differential
rotation may pose the most difficult problem. Allowance for weakly
differential rotation is possible for small nonradial
perturbations~[13, 16]. However, nonaxisymmetric (in the limiting
case, spoke-like) perturbations are more unstable. Their
stabilization under the same conditions requires significantly
stronger disk heating, i.e., a higher stellar velocity
dispersion~[17], and these very perturbations apparently increase
the stellar velocity dispersion at the nonlinear stage in the
case of initially cool systems.

Analytical studies of the gravitational stability of disks usually
apply the epicyclic approximation $c_r \ll V$, which is valid at
the peripheries of most galaxies, where $c_r/V \simeq 0.1\div
0.3$, but breaks down near their centers, where $c_r \gee V$.

Another important limitation of analytical approaches is that the
criteria derived are local, whereas a number of studies suggest
that the disk stability conditions have a global nature~[18, 19].
This means that  the equilibrium parameters, e.g., at the center,
while leaving their values at the disk periphery unchanged, may
affect the conditions for gravitational stability throughout the
disk. In contrast to a local approach based on the analysis of
dispersion equations, global analyses aim to determine the
eigenfrequency for the entire disk by solving a boundary-value
problem, which determines the influence of the conditions in one
part of the system on the dispersion properties of the
perturbations throughout the disk. Therefore, in a rigorous
approach, the disk must be considered as a whole. However, global
analyses have been performed only for certain specific power-law
distributions~[18--20]. For example, Bertin \textit{et~al}.~[21]
considered global modes in galactic disks as possible mechanisms
for maintaining long-lived spiral density waves.

Numerical simulations of collisionless systems are more flexible
in terms of the choice of model. They make it possible to go beyond
simple two-dimensional models and directly follow the development
of perturbations in a disk that is initially in equilibrium. However,
this approach has drawbacks of its own. The most serious problems
of $N$-body simulations include (1) certain mathematical simplifications
that are inevitable when a disk is modeled as a system of $ N$
gravitating bodies, where $N$ is incomparably smaller than the number of
stars in real galaxies, and (2) the dependence of the final state
of the system (after it reaches quasi-equilibrium) on the initial
parameters, which are poorly  known for real galaxies. When comparing
simulation results with real galaxies, it can also be difficult to
allow for the dissipative galactic medium (gas), in which the
sound speed is much lower than the stellar velocity dispersion.
We consider these problems in more detail below.

The principal goal of this study is to determine for galaxies with
various mass distributions the minimum local disk velocity
dispersions that enable their three-dimensional disks, which are
initially in a weakly unstable equilibrium, to reach a
quasi-stationary state. Sections~2 and 3 give a concise review of
previous results of analytical and numerical approaches to
estimating the stellar velocity dispersions required to ensure
the stability of collisionless disks. Section~4 describes the
principles underlying the construction of the dynamical models
and the determination of the stability threshold used in this
paper. Section~5 considers various three-dimensional models of
galaxies, and the last section presents and discusses the main
results.

\vfill\eject

\section{ANALYTICAL STABILITY CRITERIA}

Several criteria for gravitational instability derived analytically
using various models have been published. Let us review those we
consider to be most important.

\noindent
\textbf{(a) The Toomre Criterion}

In order for an infinitely thin, uniform, rigidly rotating stellar disk to
be gravitationally stable against axisymmetric perturbations, it
must satisfy the following condition derived by Toomre~[1]:
\begin{equation}\label{QToomre:Khoperskov_n}
 c_r \geq c_T={3.36 G \sigma \over \varkappa} , \qquad Q_T={c_r\over c_T} \geq 1,
\end{equation}
 where $\varkappa=2\Omega\sqrt{1+rd\Omega/2\Omega dr}$ is the
epicyclic frequency and $\sigma$ is the surface density.
Condition~(1) assumes that the epicyclic approximation is valid,
i.e., that the difference between the velocity $V(r)$ and the
circular velocity $V_c(r)$ can be neglected. Although this
relation was derived in a local analysis, the study of Evans and
Read~[19] of the eigenmodes for self-similar disks in a
corresponding approximation overall supports its validity.
Miller~[22] compared the theoretical increments derived for a
Toomre model with the results of simulations of axisymmetric modes
in which all other perturbations were artificially suppressed.
Experimental gradients were shown to be consistent with the
theoretical results.

\vspace*{5pt}
 \noindent \textbf{(b) Allowance
for Finite Disk Thickness}

Finite thickness is a stabilizing factor for gravitational
instability in the plane of the disk~[9, 12]. The generalization of the
Toomre stability criterion~(1) for the case of a disk of finite
thickness obtained by Morozov~[13, 14, 16] has the form
\begin{equation}\label{QToomreVander:Khoperskov_n}
\label{} Q_T^{(1)} = {1\over 1+0.974 \Delta\varkappa/c_r}  < 1 ,
\end{equation}
where $\Delta$ is the half-thickness of the isothermal
self-gravitating disk. However, this condition proved to be far
from sufficient to ensure the stability of real systems. This
became clear from the first numerical simulations performed in the
1970s, which showed that $c_r \simeq (1.5\div 5)\times c_T$ at the
periphery of a stationary, collisionless disk~[23--30].

\vspace*{5pt}
 \noindent \textbf{ (c) Simplified Allowance for
Nonaxisymmetric Perturbations}

The stronger instability of spiral waves compared to
axisymmetric modes is one of the factors that makes the Toomre
criterion inadequate. Polyachenko and Shukhman~[31],
Kalnajs~[32], and Polyachenko and Fridman~[9] were the first to
show that nonaxisymmetric modes are the dominant instabilities in
a gravitationally unstable disk. As pointed out by Polyachenko and
Fridman~[9], the azimuthal-velocity dispersion $c_\varphi$ is
smaller than $c_r$ (except for the innermost regions). Therefore,
the relation
\begin{equation}\label{QkappaOmega:Khoperskov_n}
c_\varphi = c_r  {\varkappa \over 2 \Omega},
\end{equation}
implies that, in order for an azimuthally cooler disk to be
stabilized, it must be heated more strongly, so that, in view
of~(3), the condition~(1) can be rewritten in the form
\begin{equation}\label{QTDelta:Khoperskov_n}
\label{} Q_T^{(2)} \geq S , \qquad  {\rm where} \quad S =
\frac{2\Omega}{\varkappa}.
\end{equation}
The parameter $S$ characterizes the degree of differential
rotation of the disk. In most cases, we can assume that $1\leq
S\leq 2$, based on the observed shapes of galactic rotation
curves. This form of the stability condition is discussed
in~[13--16].

Thus, the azimuthal velocity dispersion determines the elasticity
of the medium against strongly nonaxisymmetric perturbations, so
that the suppression of gravitational instability requires
stronger heating of the disk by a factor of $2\Omega/\varkappa$.

Criterion~(4) can be considered to be the Toomre criterion with a
simplified allowance for nonaxisymmetric perturbations.

\noindent
\textbf{(d) The Morozov Criterion}

Morozov~[13], Morozov and Khoperskov~[14], and Morozov~[16]
analyzed the dynamics of weakly nonradial perturbations in a nonuniform
disk in a WKB approximation. The resulting stability criterion takes
into account many factors (radial nonuniformity of the surface density
$\sigma$ and radial-velocity dispersion $c_r$, the disk thickness, and
differential rotation):
\begin{equation}\label{QMorozov}
Q_T^{(M)}=c^{(M)}/c_T, \,\,\, c^{(M)} = S\,D\,c_T\,\left\{1 +
1.07\cdot \left\vert 1.87\,S \,{c_T \over  \varkappa}\,\left({d
\sigma \over  \sigma dr} + {dS\over  Sdr} - 1.09\,{d c^{(M)} \over
c^{(M)} dr} \right)\right\vert^{2/3}\right\},
\end{equation}
where $S = 2\Omega/\varkappa$ and $D = (1 +
0.974\varkappa\Delta/(S^2 c_T))^{-1}$.

 In the general case,
determining $c^{(M)}$ from~(5) amounts to integrating a reduced
differential equation~[33]. The main drawback of this criterion is
that it was derived in the context of the dynamics of tightly
wound spiral waves ($m/r \ll k_r$, where $k_r$ is the radial
wavenumber) and then formally applied to spokelike perturbations.

If the scale length $L_c \equiv (d\ln c^{(M)}/dr)^{-1}$ is fixed,
the differential equation~(5) reduces to a simple algebraic
relation~[13, 14, 16]. For typical rotation-curve shapes, we have
$ D\simeq 0.6\div 0.8$. Introducing the factor $D$ also makes it
possible to formally allow for the disk thickness in criteria~(3)
and (6).

\noindent
\textbf{(e) The Polyachenko--Polyachenko--Strel'nikov Criterion}

Unlike the criteria discussed above, the analysis of Polyachenko
\textit{et~al}.~[17] focused on extremely nonaxisymmetric
perturbations in a thin disk. Under the assumptions made, the
rotation curve is the sole factor determining the stability
boundary, since this boundary depends only on the parameter
${n\equiv -{rd\Omega/(\Omega dr)}}$. Figure~1 in~[17] shows the
dependence of the dimensionless velocity dispersion at the
stability boundary on the parameter $\alpha = 2/n$. We will use
the approximating function
\begin{equation}\label{QTPol}
 Q_T^{(P)}={c^{(P)}\over  c_T} = 1.88\, \sqrt{ 1.1 + {8 \over  \exp(\alpha -
1/4) - 1} }\,,
\end{equation}
where $Q_T^{(P)}$ is the minimum Toomre parameter for a stable
disk.

This approximation has sufficient accuracy for our needs when
$1.2< \alpha^2$. The criterion of Polyachenko \textit{et~al}.~[17]
depends on a single parameter, since the form of the rotation
curve fully determines the stability boundary. In particular, it
follows that $Q_T^{(P)}\simeq 3$ in the region with constant
rotational velocity ($n = 1$).

\noindent
\textbf{(f) Allowance for the Gaseous Subsystem in a Stability Analysis
for Stellar Disks}

The presence of a cooler component also contributes to
destabilization of the stellar disk. A number of authors~[34--38]
have analyzed this problem in detail as applied to radial
perturbations. Ortega \textit{et~al}.~[38] considered a more
general problem: they analyzed how the inhomogeneous composition
of a thin disk influences the stability of the disk against small
radial perturbations when the disk consists of particles having a
mass spectrum such that more massive particles have lower velocity
dispersions.

The problem is simplified if the mass of the ``cool'' component is
relatively small. Assuming the gas surface density
$\sigma_{\textrm{gas}}$ to be usually much lower than the surface
density of the stellar disk $\sigma_{\textrm{star}}$ and ${c_s \ll
c_r}$ (where $c_s$ is the adiabatic sound speed in the gas),
onecan write for the velocity dispersion of a stellar disk
containing gas at the stability boundary~[37]
\begin{equation}\label{star-gas-crit}
{c_r^{crit} \over  c_T} = 1+{\sigma_{gas} \over
\sigma_{gas}+\sigma_{star}} {1-(c_s / c_T)^2 \over  1+(c_s /
c_T)^2}
 \,.
\end{equation}
For example, for the parameters of the solar neighborhood in the
Galaxy, it follows from~(7) that the gaseous subsystem, which
contributes about $20\%$ of the disk surface density, increases
the minimum stellar-velocity dispersion sufficient for stabilizing
radial perturbations by $\leq 10\%$.

Finally, we note that attempts have also been made to determine
the parameters of the disk subsystem using other approaches based
on the possible existence of various structures in the disk
(spiral density waves or a bar) rather than on local stability
conditions, assuming certain mechanisms that form and sustain
these structures. However, the analysis of these approaches is
beyond the scope of this paper.

\section{ANALYSIS OF GRAVITATIONAL STABILITY  IN NUMERICAL SIMULATIONS}

Numerical simulations describing the dynamical evolution of disks
can be used to analyze disk instabilities for given initial
conditions allowing for nonuniform distributions of the mass and
angular velocity. Such simulations naturally take into account the
formation of appreciably nonlinear and nonaxisymmetric structures
such as bars or transient spirals. The main problem with
this approach is that the results depend on the initial conditions,
since the evolution of a disk starting from a strongly unstable state
can proceed very differently from the evolution of a disk starting in a
subcritical state. In addition, the disk can
suffer from slow secular instabilities that are difficult to take
into account in numerical simulations.

$N$-body evolutionary models are usually aimed at studying the
development of instability or --- which is of the most interest
for us --- at analyzing the states of a system after many disk
rotations. Of the large number of published studies of this type,
we will mention those that are, in our view, most important in
the context of establishing the conditions for stability of a
disk.

Carlberg and Sellwood~[39, ~40] analyzed the influence of small,
nonstationary perturbations of the potential on the evolution of the
velocity-distribution function. In particular, they analyzed scattering by
nonstationary spiral waves; the resulting growth of the stellar-velocity
dispersion agrees well with the results of numerical simulations.

The critical velocity dispersion for stellar disks (the parameter
$Q_T$) has been computed many times based on the results of
$N$-body dynamical simulations (see, e.g.,~[7, 24, 28, 33,
40--46]). None of the simulations yielded a stable disk with $Q_T
< 1$. As a rule, $Q_T(r)$ increases with distance from the center
of a galaxy. This pattern appears both in simulations of
three-dimensional disks~[5, 46--48] and in some theoretical
analyses~[49].

The work of Athanassoula and Sellwood~[34] can be considered to be
classic. They concluded that a two-dimensional disk is always
stable if its Toomre parameter exceeds $Q_T \gee 2.2\div 2.4$.
However, this conclusion did not take into account vertical
motions. Since three-dimensional disks are gravitationally more
stable, two-dimensional models underestimate the mass of a
marginally stable disk. In addition, Athanassoula and
Sellwood~[34] used a radially averaged $Q_T$ and simulated a
specific density distribution based on a Toomre--Kuz'min model. It
is therefore not clear whether their results are applicable for
real three-dimensional disks.

\section{NUMERICAL MODELS:  SPECIFYING INITIAL
CONDITIONS}

The dynamical models used in this paper are based on numerical
integration of the equations of motion for $N$ gravitationally
interacting particles taking into account the external field
produced by the steady-state mass distribution in the bulge and halo.

For the halo, we adopted the commonly used spatial distribution of
the volume density in the form
\begin{equation}\label{densityhalo}
\varrho_h(r) = {\varrho_{h0}\over  (1+\xi^2/a^2)^k} \,,
\end{equation}
where $\xi=\sqrt{r^2+z^2}$ is the radial coordinate. Choosing
$k=1$ yields a flat rotation curve in the halo-dominated region.
The relative central density of the spherical component increases
with $k$, imitating the influence of the bulge. Because the bulge
enters our model as a separate spherical component, we restrict
our analysis to a halo model with $k=1$.

The following set of equations describes the dynamics of the $N$
gravitating bodies:
\begin{equation}\label{LawofNewton}
{d^2\vec{r_i}\over  dt^2} = \sum_{j}^{N}{\vec{f}_{ij}} + \vec{F}_s
\,\,\,\,\,\,\, (i=1,...,N)\,,
\end{equation}
Here, the radius vector $\vec{r}_i(t)$ determines the position of
the $i$th particle, $\vec{f}_{ij}$ is the force of interaction
between the $i$th and $j$th particles, and the force
$\vec{F}_s=\vec{F}_b + \vec{F}_h$ is due to the bulge/halo
spheroidal subsystem. The halo mass distribution~(8) with a
central density of $\varrho_{h0} = M_h/\{ 4\pi a^3 [R/a
-{\tan}(R/a)]\}$ yields for the force
\begin{equation}\label{Forcehalo}
\vec{F}_h(\vec{r}) = - {4\pi G a^3 \varrho_{h0} \over
\xi^2}\left\{ {\xi\over  a} - {\rm arctg}\left({\xi\over
a}\right)\right\}\, {\vec{r}\over  \xi} \,
\end{equation}
which is determined by the spatial scale lengths $a$ and mass
$M_h$ inside the sphere $\xi=|\mathbf{r}|<R$. We adopt a King
model for the density distribution in the spherical bulge:
\begin{equation}\label{densitybulge}
\varrho_b=\left\{
\begin{array}{cl}
\varrho_{b0}/\left[1+(\xi/b)^2\right]^{3/2}&,\ \ \xi <(r_b)_{\max}\\
0&,\ \ \xi >(r_b)_{\max}\,,\\
\end{array}
\right.
\end{equation}
where
\begin{equation}\label{massbulge}
\Oo M_b = 4\pi b^3 \varrho_{b0} \bigg\{ \ln \Big[ (r_b)_{\max}/b +
\sqrt{1+((r_b)_{\max}/b)^2} \Big] - {(r_b)_{\max}/b \over
\sqrt{1+((r_b)_{\max}/b)^2}} \bigg\}
\end{equation}
is the mass of the bulge. We have for the gravitational force
inside $\xi \le (r_{b})_{\textrm{max}}$
\begin{equation}\label{Forcebulge}
\vec{F}_b = - {4\pi G b^3 \varrho_{b0}\over  \xi^2} \bigg\{
\ln\bigg( {\xi\over  b} + \sqrt{1+{\xi^2\over  b^2}} \bigg) -
{\xi/b \over  \sqrt{1 + \xi^2/b^2}} \bigg\} \, {\vec{r}\over  \xi}
\,.
\end{equation}
It is obvious that $\mathbf{F}_b = - GM_b \mathbf{r}/\xi^3$ in the
domain $r > (r_b)_{\textrm{max}}$.

The dynamical model must adequately describe the Newtonian
interactions between stars and ensure that the system is
collisionless. This is achieved by modifying the gravitational
force at small distances by introducing a potential-cutoff radius
$r_c$ for any pair of interacting particles $i$ and $j$. The
optimum choice of cutoff radius and number of particles has been
widely discussed in the literature (see, e.g.,~[50--53] and
references therein).

Here, we use a Plummer model for the potential:
\begin{equation}
\label{ModelPlummer} \Phi_{ij} = - G{m_i m_j\over  \sqrt{r_{ij}^2 +
r_c^2}} \,,
\end{equation}
where $r_{ij}$ is the distance between particles and $r_c$ is the
cutoff radius. For a fixed number of particles $N$, it is always
possible to choose a cutoff radius $r_c$ that ensures that the
model is collisionless. However, the number of particles must be
sufficiently large to minimize the error introduced by the
modification of the particle-interaction potential at small
distances.

We characterize the disk surface density by the scale length $L$,
which determines the exponential law $\sigma(r) = \sigma_0
\exp(-r/L)$. We assumed that the disk surface density is zero in
the region $r\geq 5L$ at the beginning of the simulations. We used
a system of units in which $G=1$, $R=4L=1$, and the mass of the
disk is $M_{\textrm{d}}=1$.  We normalized the mass of the halo
$M_h$ inside the radius $\xi \leq 4L$ to the mass of the disk,
$\mu\equiv M_h/M_{\textrm{d}}$. In this system of units, one
period of rotation of the outer edge of the disk lies in the range
$t\sim 3\div 4$ for $\mu=1\div 4$.

Let us now describe the procedure for specifying the initial
density distribution along  $z$ axis and the residual velocities
of the equilibrium disk.

The vertical structure of the disk is determined by the equations~[54]
\begin{equation}\label{EquationBahcall}
{1\over  r}{\partial \over  \partial r}\left( {r{\partial \Phi
\over \partial r }}\right)+{\partial ^2 \Phi \over  \partial
z^2}= 4\pi G(\varrho + \varrho_s) \,, \ \ \ \ \ \ c_z^2
{\partial  \varrho \over
\partial  z} = - {\partial  \Phi \over  \partial  z}\, \varrho     \,,
\end{equation}
where $\varrho$ and $\varrho_s$ are the spatial density in the
disk and spheroidal subsystems, respectively, and $c_z$ is the
vertical-velocity dispersion, which is assumed to remain constant
with $z$ at $t=0$. Eliminating the potential $\Phi$ from~(15) and
introducing the circular velocity
\begin{equation}\label{Vc}
\Oo V_c(r) \equiv \sqrt{ r \left( {\partial  \Phi \over
\partial  r} \right)_{\big| z=0}} \,,
\end{equation}
we can transform the set of partial differential equations to an
approximate equation for the disk volume density $\varrho(z)$ in
the form of an ordinary differential equation ~[54]:
\begin{equation}\label{Bachall}
\varrho{d\over  dz}\left( c_z^2 {d\varrho\over  dz}\right) - c_z^2
\left( {d\varrho\over  dz} \right)^2 + 4\pi G \varrho^2 \big(
\varrho + E+\varrho_s(z)\big) = 0 \,, \ \ \ E=-{1\over  4\pi G
r}{dV^2_{c}\over  dr}\,.
\end{equation}
Together with the conditions $\varrho(z=0) = \varrho_0$,
$d\varrho(0)/dz=0$, and
$\int\limits_{-\infty}^{\infty}{\varrho(z;r)dz}=\sigma(r)$, this
equation determines the vertical structure of the disk at the
radius $r$ for a given surface-density distribution $\sigma$. The
$E$ term can produce a large error in the estimated density at the
disk center, and, in practice, it is assumed that ${E(r\rightarrow
0)\rightarrow 0}$. To determine $\varrho_0$ and $\varrho(z)$ for
specified $\varrho_s(z,r)$, $c_z(z,r)$, and $\sigma(r)$, we
construct the function
$F(\varrho_0)=2\int\limits_0^{\infty}{\varrho(z)dz}-\sigma$. We
solve the equation ${F(\varrho_0)=0}$ iteratively jointly with
numerical integration~(17). After determining the density
distribution in $z$, the particles are arranged along the vertical
axis on a   grid $z_k = k \Delta z$ (${k=-K,\ldots, K}$). The
$k$th cell contains particles in proportion to $\sigma_k/\sigma$,
where $\sigma_k=\int\limits_{z_{k-1}}^{z_k} \varrho(z) dz$.

Equation~(17)  is approximate, since it was derived neglecting the
dependence of the potential on the vertical coordinate in the first
term in~(15). Strictly speaking, a disk constructed in this way is not
in equilibrium. However, we are interested in initial, unstable states
that evolve to new stationary states of the disk. Therefore, the lack
of an exact equilibrium in the vertical direction plays the role of a
small additional initial perturbation.

The initial velocity distribution is a Schwarzschild function and
has the form of an anisotropic Maxwel\-lian distribution:
$$
f(u, v, w) = A\, \exp\left\{ -{u^2 \over  2\,c^2_r} -
{(v-r\Omega)^2 \over  2\,c^2_\varphi} - {w^2 \over
2\,c^2_z}\right\}\,,
$$
where $\{u,v,w\}$ are the velocity components of the particles in
cylindrical coordinates.

To obtain the model with the minimum velocity dispersion in the
final, stable disk state, we chose a subcritical disk state whose
dynamical evolution ensured  both the preservation of an
exponential surface-density profile and stability against initial
perturbations in the plane of the disk and against bending
perturbations (responsible for the increase in the vertical
velocity dispersion) at the end of the computations.

In practice, the initial radial-velocity dispersions $c_r$ in
models with low-mass bulges corresponded to Toomre parameters
${Q_T\simeq 0.8\div 1.1}$ and ${Q_T\simeq 1.2\div 2.2}$ in the
central region ($r\lee 2L$) and at edge of the disk, respectively.
In the case of massive bulges, the models started from higher
central values of $Q_T$. The initial vertical-velocity dispersion
was set $c_z$ proportional to $c_r$. We considered two types of
models with different initial disk thicknesses: ``thin'' disks
unstable in the $z$ direction\footnote{The bending instabilities
of the oscillation mode lead to the heating of a thin disk in the
vertical direction and, as a result, to its thickening. As our
numerical models show, the axisymmetric bending oscillation mode
plays an important role.}, which had central $(c_z/c_r)_0$ values
$(c_z/c_r)_0 = (0.4\div 0.5)$ at $t=0$, and ``thick'' disks with
$(c_z/c_r)_0 = (0.6\div 0.8)$, which are close to the stability
limit against bending perturbations. We assumed that these ratios
varied slowly in radius in accordance with an exponential law with
a radial scale length appreciably exceeding $L$. This choice of
velocity dispersion ensured weak instability of the disk at all
$r$. With the exception of specially stipulated cases, we describe
below numerical models of ``thick'' disks.

In none of the models did the velocity dispersion $c_r$ remain
constant: the disk ``heated up'' in the course of its evolution.

We found the mean tangential velocity of the model point masses
by solving the Jeans equation assuming the absence of systematic
radial motions, axial symmetry, and symmetry about the plane $z=0$:
\begin{equation}\label{VelocityRotation}
V^2=(< v >)^2 = V_c^2 + c_r^2\, \left\{ 1 - {c_\varphi^2\over
c_r^2} + {r\over  \varrho c_r^2}{ \partial  (\varrho c_r^2)
\over  \partial  r} + {r\over  c_r^2}{\partial  <{uw}> \over
\partial z} \right\} \,,
\end{equation}
Here, $<\ldots>$ denotes averaging of the velocities and the last
term in~(18) is due to the chaotic components of the radial
velocity $u$ and vertical velocity $w$. When specifying the
initial state of the model, we assumed that $<u>=0$ and $<w>=0$
and assigned the rotational velocity of the disk in accordance
with~(18). Thus, initially we have a balance of the radial and
vertical forces, and hence the disk begins to evolve from a nearly
equilibrium state.

In all the computations, we specified the initial distribution of
the velocity dispersion $c_\varphi$ at $t=0$ in accordance
with~(3). We verified that the condition $\Oo{Q_c\equiv {c_r\over
c_\varphi}{\varkappa\over 2\Omega}=1}$ was satisfied as the disk
evolved. In models with fairly massive halos ($\mu\gee 2$), the
mean deviations of $Q_c$ from unit at a given $r$ did not exceed
$3\%$ over several dozen rotations of the disk edge. This error is
partially due to the numerical differentiation used in a process
of  computing the epicyclic frequency $\varkappa$. In the models
with massive halos, the domain in which ${c_r / \varkappa} > r$ is
small (for $\mu=3$, we have  $r\lee 0.03$). In the case of
$\mu\lee 1$, this domain expands to $r\lee 0.15$. In addition, the
vertical disk scale length increases in such models, and these
factors result in stronger deviations from the equality~(3). The
amplitude of fluctuations $Q_c(t)$ decreases as the number of
particles increases  and does not exceed $2\%$ in models with
$\mu\gee 2$ and $N=2\times 10^{5}$ (except for the innermost
region of the disk, $r\lee 0.5L$), which reflects the
\emph{collisionless} nature of the models constructed.

Using the approach described above, we performed more than 40
numerical simulations of the dynamical evolution of a disk toward
a steady state in dependence of the initial velocity dispersions
$c_r(r)$ and $c_z(r)$. We considered a wide range of parameters
for the bulge ($M_b/M_{\textrm{d}}=0\div 3$, $b/L=0.02\div 0.8$)
and halo ($M_h/M_d=0\div 3.5$, $a/L=1\div 4$). The number of
particles was $N=(40\div 500)\times 10^3$ in the TREEcode
computations.

We verified the stability of the solutions against the choice of
computational method by comparing results for several models
obtained using two very different methods to compute the
gravitational force: the TREEcode method and direct ``particle to
particle'' (PP) integration, in which each particle interacts with
each of the other particles, for $N=(20\div 80)\times 10^3$. As an
example, Figure~1 shows the time dependence of the radial-velocity
dispersions for one of the models computed using these two
methods, with similar initial conditions.\footnote{The
velocity-dispersion components in the plane of the disk $c_r$ and
$c_\varphi$ increase rapidly during the initial stage ($t<10$) of
the heating of an initially cool disk with initial Toomre
parameter $Q_T(r<2L) \simeq 0.85$. The relaxation of the disk in
the vertical direction is appreciably slower, so that small local
decreases of $c_r$ are possible when the disk is heated in $z$ or
there are radial motions (see curve \emph{1} in Fig.~7 after
$t>7$).} A comparison of the two results reveals no significant
differences between the final disk states.

\section{DETERMINING THE THRESHOLD $Q_T$ VALUES}

\subsection{Disk Heating Mechanism}

An initially axisymmetric, equilibrium disk is heated, increasing
its velocity dispersion with time. The question of greatest
importance for dynamical simulations is the heating mechanism. To
correctly describe the processes in stellar disks, it is
important, in particular, to ensure that the heating is not due to
the collisional relaxation of the particles, whose number $N$ is
many orders of magnitude smaller than the number of stars in real

systems. This is achieved via appropriate choices of the number of
particles and the cutoff radius. Our criterion for the absence of
significant collisional-relaxation effects is that the character
of the system's evolution be preserved when the computations are
performed for an increased number of particles, as we verified for
many models\footnote{The minimum number of particles
$N_{\textrm{crit}}$ for which this criterion is satisfied to
sufficient accuracy (i.e., to within random fluctuations of the
estimates of the final parameters) depends, in particular, on the
form of the rotation curve. The number $N_{\textrm{crit}}$ is
larger for highly concentrated nuclei (bulges), due to the higher
rotational angular-velocity gradient in the central region of the
disk. In the limiting case of a very short, rigidly rotating
region, $N_{\textrm{crit}}$ can reach $3\times 10^6$~[20]. Our
simulations showed that, if the rotation curve grows monotonically
to $r \simeq 2L$, then $N_{\textrm{crit}}$ can be set equal to
$\simeq 4\times 10^4$. }.

Below, we summarize the most important features of the disk
heating shown by the numerical simulations.

(1) The disk-heating time is much greater than the mean rotation
period of the particles (Fig.~2a). In the initial stage ($t\lee
1$), $c_r$ remains virtually constant, as long as the disk remains
axisymmetric. In the case of a low-mass or nonexistent halo, the
evolution of the disk is determined by the bar mode, and the disk
is heated due to the formation of an nonaxisymmetric bar and
associated two-armed spiral. Models with sufficiently massive
haloes do not show any enhancement of the bar mode,however they
develop a complex transient system of small-scale spiral waves
(Fig.~2b). The decrease of the amplitudes of these waves with time
is accompanied by the transfer of rotational kinetic energy to the
chaotic component of the velocity, resulting in heating of the
disk.

(2) The heating of an initially cool disk ($0.5\lee Q_T\lee 1$)
begins in its central region (Figs.~2a, 2b). The heating at the
periphery proceeds much more slowly. The velocity dispersion at
the periphery usually begins to rise when the center has already
reached a quasi-stationary state (Fig.~2a). At the same time, the
processes at the center  and at the periphery are interrelated:
fast growth of instability in the central region can speed up the
heating of the outer part of the disk, while stability of the
central region can slow down this process.

3) If the system does not develop a bar, the amplitude of the
perturbations begins to decrease with increasing velocity
dispersion. In turn, the increase of the radial-velocity
dispersion $c_r$ slows down parallel with decreasing wave
amplitude. The heating virtually ceases after the decay of the
transient spiral waves (see Figs.~2a, 2b). The integrated
amplitude of the Fourier harmo\-nics
 $$\hat{A}(m;t)=\sqrt{\sum\limits_{p} A^2(m,p;t)}\,,\ A(m,p,t) =
{1\over  N}\  \sum\limits_{j=1}^{N}\ \exp\left\{ i \left[
m\varphi_j(t) + p \ln (r_j(t)) \right] \right\}
 $$
decreases with
time for all mode numbers $m$, but most slowly for $m=2$
(Fig.~2c). The density distribution in the disk becomes almost
axisymmetric (if the mass of the spherical components is large
enough to prevent the development of a bar), and, on the whole,
the velocity dispersion $c_r$ maintains its level over several
dozen rotations, provided that relaxation processes have ceased in
the vertical direction.

(4) Whereas three- and even four-armed modes can dominate at the
initial stage of evolution of an initially cool disk ($Q_T(r\le
2L)< 1$; see Figs.~2b, 2c), the spiral pattern changes if we
choose a subcritical initial state (i.e., a state that is
unstable, but not so cool, where $Q_T\gee 1)$. The spatial
structure of the perturbations depends to a considerable degree on
the relative mass of the spheroidal subsystem. Figure~3 shows the
distributions of the logarithm of the disk surface density at
various times. The two-armed mode is dominant in this model,
although the $m=3$ harmonic is also important, especially in
initial stages of the evolution. It is typical for the spirals to
join in the outer region of the disk to form a ring-shaped
structure.

(5) Models with low-mass spheroidal subsystems starting from a
very cool initial state undergo a substantial mass redistribution
in the disk, accompanied by a change in the form of the rotation
curve $V(r)$ in the process of heating and formation of a bar. In
this case, the final distribution of the surface density
$\sigma(r)$ differs strongly from an exponential law
(Fig.~4).\footnote{It is possible, in principle, to choose an
initial density profile that yields an exponential profile at the
end of the computations (dashed line in Fig.~4); however, this
approach is rather artificial. It seems that the stellar disks of
most of the galaxies do not pass through a stage of strong
dynamical instability.}

Another feature of model disks starting from a very cool initial
state ($Q_T\lee 1)$ is that the velocity dispersion at the end of
the computations (after 10--15 rotations) is somewhat higher than
is required for gravitational stability. This is due to heating by
collective processes---high-amplitude wave motions that develop in
the presence of strong instability. When the disk is heated and
reaches marginal stability, these perturbations die out, but the
wave-decay process has a certain inertia: the velocity dispersion
is already high enough to maintain the stability of the disk, but
the spiral waves have not yet decayed (as is confirmed by Fourier
analysis of the density perturbations in the disk) and continue to
heat the disk. Therefore, to obtain the minimum velocity
dispersion required for disk stability, we used an iterative
algorithm to make the initial velocity dispersion approach the
stability limit.

Our \emph{iterative approach} is based on a series of several (two to four)
consecutive computations, each of which starts with an initial dispersion
that is somewhat closer to the critical level than that for the previous
computation. For each radius $r$, we chose an initial velocity dispersion
that was intermediate between the initial and final values (after five to
ten rotations) in the previous computation.

As expected, the stability limit also depends on the initial disk
thickness\footnote{The initial thicknesses of the disks of real
galaxies depend on the conditions under which they were formed.
For example, if the collapse of a gaseous disk was accompanied by
violent star formation before it reached a quasi-stationary state,
the resulting stellar disk may be much thicker and have a higher
velocity dispersion $c_z$ than the minimum value required for
stability. However, such a scenario is difficult to reconcile with
the existence of very thin ($h_z/L <  0.2$) stellar disks in
edge-on galaxies. The dependence of the disk thickness on the
relative mass of the halo is also consistent with the assumption
that the stellar-velocity dispersion is close to the value
expected for marginal stability~[7].}. If the disk is initially
thick (with the vertical scale height $h_z\gee 0.2L$) and unstable
only in its plane ($c_r < c_r^{\textrm{crit}}$), its heating
proceeds more slowly and ceases at lower radial-velocity
dispersions than in the case of an initially thin disk. This is
due to two factors: the stabilizing effect of the finite thickness
of the disk and the slowness of the relaxation in the vertical
direction compared to the heating in the plane of the disk.

The results of dynamical simulations allow to determine the disk
parameters at the gravitational-stability limit (when the velocity
dispersion ceases to change, after $5 \div 20$ rotations of the
outer edge of the disk).

\subsection{Gravitational-Stability Threshold for Bulgeless Models}

We computed a series of bulgeless galaxy models, differing in the
relative mass $\mu=M_h/M_{\textrm{d}}$ and the halo scale length
$a$, with a fixed radial disk scale~$L$.

When $a\gee L$, the rotation curve has an extended section
$V_c(r)$ (which we will call the ``rigid-rotation'' section),
which makes a transition to a plateau $V_c\simeq const$ at $r\gee
2L$ (curve~\textit{1} in Fig.~5a). Figure~5a shows for a bulgeless
galaxy with $\mu=1$ the radial distributions of the circular
rotational velocity $V_c(r)$ (curve~\textit{1}), mean particle
rotational velocity $V(r)$ (curve~\textit{2}), radial-velocity
dispersion $c_r(r)$ (curve~\textit{3}), and the
differential-rotation parameters $S = {2\Omega /\varkappa}$
computed separately for $V_c(r)$ and $V(r)$ (4 and 5).

Knowing the final velocity dispersions, disk density, and rotational velocity,
it is possible to compare the Toomre stability parameters $Q_T$~(1) for
the corresponding model with the analytically derived local criteria
discussed above and compare these criteria with each other. Note that the
analytical criteria were derived assuming that the circular rotational
velocity $V_c(r)$ differs little from the mean velocity $V(r)$ of the
gravitating point masses. However, the difference between these velocities
can be quite significant for collisionless disks. We therefore determined the
stability parameters for both $V_c(r)$ and $V(r)$.

Figure~5b shows for this model the computed $Q_T$ and the Toomre
parameters required for marginal stability according to the
criteria of Polyachenko, Polyachenko, and Strel'nikov (PPS)
$Q^{(P)}_{T}$~(6) and Morozov $Q_T^{(M)}$~(5) calculated
separately using the circular velocity and the mean rotational
velocity of the particles. In the case of a massive halo ($\mu
\gee 2$), the difference between the stability parameters for
$V_c(r)$ and $V(r)$ is small, but it can become appreciable for
$\mu \lee 1$ (Figs.~5a, 5b). This is due to the fact that the
difference $V_c-V$ increases with increasing velocity dispersion
in accordance with~(18).

The condition ${S \simeq Q_T^{(M)} \lee Q_T \lee Q_T^{(P)}}$ is
satisfied both in the case considered here and for most of the
models (recall that $Q_T=S$ is the Toomre condition for marginal
stability with crude allowance for nonaxisymmetric perturbations;
see Section~2).

Figures~5c and 5d compare the stability criteria in a different
way. These figures show the radial dependences of the ratio of the
model velocity dispersion $c_r$ of the disk when it becomes stable
to the critical velocity dispersion for the various criteria
considered in Section~2. A perfect agreement with the model

dependences would correspond to the ratio $c_r/c_{\textrm{crit}}$
= 1.

Although none of the criteria explain the model dependences
$Q_T(r)$ at all $r$, the closest results are given by the PPS
criterion generalized for the case of finite thickness~(2) and the
Morozov criterion~(5) (symbols 15 and 16 in Fig.~5c). When the
stability criteria are determined in terms of the mean rotational
velocity (Fig.~5b), the radial variations of
$c_r/c_{\textrm{crit}}$ conserve their general form (Fig.~5d).
Note that for the interval $0.1<r<0.8$ the ratio
$c_r/c_{\textrm{crit}}$ is closest to unity for the Morozov
criterion (curve 16). It is striking that the PPS criterion
(curve 12) overestimates $c_{\textrm{crit}}$ in all cases, but
$c_r/c_{\textrm{crit}}$ remains nearly constant over a wide range
of $r$, enabling easy introduction of a correction factor for
$c_{\textrm{crit}}$.

The important result is that the radial dependence of the Toomre
parameter $Q_T(r)$ computed for the circular velocity $V_c(r)$
shows qualitatively similar behavior for all the bulgeless models
we considered (Fig.~5e). Moreover, $Q_T(r)$ remains approximately
constant at the level $Q_T\simeq 1.2\div 1.6$ in the interval
$0\lee r/L\lee 2$ (see Fig.~5e). At the disk periphery ($r\gee
2L$), $Q_T$ increases with radius, reaching $Q_T\simeq 2.5\div 3$
at the edge of the disk ($r\simeq 4L$). However, the scatter of
the $Q_T$ values for the various models is small, enabling us to
adopt the following function as a lower envelope
\begin{equation}
\label{QTour} Q_T^{(*)} = A_0+A_1\cdot \left({r\over  L}\right)
+A_2\cdot \left({r\over  L}\right)^2 \,,
\end{equation}
where $A_0=1.25$, $A_1=-0.19$, and $A_2=0.134$ (bold curve in
Fig.~5e). This function reaches its minimum, equal to 1.2, at
$r/L=0.7$.

The observational estimates of the stellar-velocity dispersions
are usually based on data for the inner region of the disk $r<2L$.
Therefore, the approximate constancy of $Q_T$ in this region makes
this quantity a convenient tool for estimating the surface density
and mass $M_d$ of a disk (assuming it is stable) based on the
known rotation curve of the galaxy and the radial exponential disk
scale length $L$. The equation relating these quantities is
\begin{equation}\label{CRour}
c_r = Q_T^{(*)}(r)\, {3.36 G \sigma_0 \exp(-r/L)\over  \varkappa}
\,,
\end{equation}
where $Q_T^{(*)}$ for galaxies with the extended region of
increasing $V_c(r)$ is determined by~(19). In turn, the disk mass
estimate $M_d=2\pi \sigma_0 L^2$ makes it possible to separate out
the mass of the dark halo within a specified radius.

\subsection{Models with Formation of a Bar}

In the case of a low-mass or very ``loose'' halo and no bulge, the
evolution of an initially cool dynamical model inevitably results
in the development of a bar. As a rule, the formation of a bar
involves an appreciable radial redistribution of the mass, whose
intensity increases when the velocity dispersion in the initial
state decreases. The outer boundary of the disk shifts outward. As
a result, the radial scale for the azimuthally averaged surface
density $L=-(d\ln \sigma/dr)^{-1}$ varies, which can lead to
deviations from an exponential profile (when $L=\textrm{const}$
throughout the disk).  In the innermost region ($r\lee 0.3L$), the
surface density can increase appreciably over its initial value
(Fig.~4), bringing about a decrease of $Q_T$ in accordance
with~(1). A similar conclusion concerning the increase in the
central density during the formation of a bar was also reached for
models with ``living'' (evolving) haloes~[55]. Numerical
simulations have shown that, under certain conditions, the
interaction between the bar and ``living'' halo can appreciably
affect the dynamical evolution of the bar~[56].

The stability criteria considered above were derived for and
applied to axisymmetric disks. However, it is of interest to
formally compute these criteria based on azimuthally averaged
parameters in the absence of axial symmetry in the inner region
--- that is for barred galaxies. As an example, Figure~6 shows
the radial distributions of the parameters of a disk that has
developed a bar (Figs.~6a--6c show results for a model with a
low-mass halo and Figs.~6d--6f results for a model with a halo
whose initial mass within $r=4L$ is twice that of the disk). The
radial distributions of the surface density in the bar region can
differ from those outside the bar (Figs.~6c, 6f). The development
of a bar usually results in additional thickening of the disk. The
function $Q_T(r)$ depends on the bar parameters (essentially, on
the initial conditions), however for $r \gee L$ the condition
${Q_{T} \gee 1.5}$ is satisfied, and the radial dependence
$Q_T(r)$ is qualitatively similar to that observed for an
axisymmetric disk. The development of a bar in models with more
massive haloes is accompanied by weaker deviations from the
initial exponential surface-density profile (Fig.~6f).

The characteristic dependences for the model with a massive halo
(Figs.~6d--6f) differ little from the cases shown in Fig.~5 for barless
models. As expected, suppressing the bar mode requires a higher initial
velocity dispersion $c_r$ or a more massive halo, in accordance with
classical concepts (see, e.g.,~[29,~44]).

\subsection{Models with Bulges}

Let us consider now model galaxies whose rotation curves in the
central region ($r\lee L$) are determined primarily by the bulge.
As a rule, the rotations curves outside the bulge $(r\gee L)$ are
almost flat ($V\simeq \textrm{const}$). Figure~7 shows typical
model radial dependences of the disk parameters for this case. In
the presence of a bulge, $Q_T$ increases strongly in the central
region of the disk, where the dynamics are determined by the bulge
potential (Figs.~7b, 7e). However, outside this region, at
$r\simeq (1\div 2)\times L$, the radial dependence of $Q_T$
retains its form [see~(19), Fig. 5e]. At larger distances $Q_T$
increases monotonically with $r$ up to values $2.5\div 3.5$. The
resulting radial dependence of $Q_T$ is typical of systems with
bulges that are not too massive ($M_b/M_{\textrm{d}} \lee 0.3$)
and not very extended ($(r_b)_{\textrm{max}} \lee L$).

The more massive and more compact the bulge, the higher the value
of $Q_T$ at the disk center. This increase of the Toomre
parameter is due mainly to the growth of the epicyclic frequency
$\varkappa$. The radial-velocity dispersion in the central region
also increases somewhat with increasing mass of a strongly
concentrated bulge. Additional
 heating of the center of the disk
($r\lee 0.5L$) in the model shown in Fig.~7a has no direct
connection with  gravitational instability. At the initial stage
of its evolution, this model, where $\mu=1$, a bar develops, which
disappears soon as a result of scattering of particles passing
near the concentrated nucleus (the latter has a scale length
$b=0.01$). This mechanism for bar disruption is similar to the
effect produced by a black hole~[57].

As a result, the disk undergoes additional heating and becomes
thicker in the central region~(Fig.~7f). The degree of this
additional heating and the circular velocity $V_c$ at the disk
center are very sensitive to the bulge parameters, first and
foremost, the core radius $b$ (in particular, the bar disruption
mentioned above does not take place if $b\gee 0.05L$). Therefore,
at $r\lee 0.5L$ the Toomre parameter $Q_T$ can vary very strongly
in different models (Fig.~7f).

\subsection{Differential Rotation as a Factor Increasing
 \protect
 the Threshold for
Gravitational Stability}

Here we consider two limiting cases: a disk that rotates almost
rigidly and a quasi-Keplerian disk whose rotational velocity
decreases as $V \propto r^{-1/2}$. Since differential rotation is
a destabilizing factor, rigidly rotating disks are expected to
have lower $Q_T$, other conditions being the same.

Galaxies usually exhibit an extended interval of almost rigid body
rotation if the mass of the disk dominates throughout most of the
disk ($r \lee 2L$), so that suppression of the bar instability
requires stronger disk heating than in the presence of massive
spherical components. Therefore, to elucidate the role of
differential rotation, we analyzed a model in which the
development of a bar is suppressed by a massive halo, yet the
outer part of the disk ($r\gee L$) contains a rigidly rotating
section. The differential-rotation parameter in this model $S(r >
L)=2\Omega/\varkappa=1.1\div 1.2$ is close to $S=1$, which
corresponds to completely rigid rotation (Fig.~8a). Figure~8a
shows the radial dependences of $Q_T$, $Q_{T}^{(P)}$, $Q_T^{(M)}$,
and $S$ characterizing the stability of the system. It is clear
that, in general, the disk becomes stable at lower values of $Q_T$
(curve $6$) than in the cases considered above. In this case we
have $Q_T\simeq 1$ and $Q_T\simeq 2$ at the disk center and
periphery, respectively.

Consider now another limiting case, corresponding to a nearly
Keplerian rotation curve (Fig.~8b). Such behavior is very rarely
seen in the rotation curves of real galaxies, but we are
interested in the fundamental question of how strong differential
rotation affects the minimum velocity dispersion required for
stability. To produce quasi-Keplerian disks, we introduced massive
concentrated components into our dynamical model. The resulting
series of models is a natural extension of those with very massive
bulges ($M_b > (2\div 4) M_{\textrm{d}}$). Figure~8b shows the
results obtained for one of such models. The Toomre parameter
increases appreciably for disks with strong differential rotation,
so that the condition $Q_T > 2$ is necessary for stability, even
in the region where $Q_T$ is minimal (${r\simeq (1\div 2)L}$).

Since the central region is dominated by the spheroidal system,
no bar develops, and there is no increase in the velocity
dispersion there, which was mentioned in Section~5.4 (Fig.~7a).

\section{DISCUSSION AND CONCLUSIONS}

We have carried out numerical simulations of the evolution of
disks that are initially unstable, collisionless, and in
equilibrium for a whole series of three-dimensional models. The
aim of our analysis was to compare the minimum radial-velocity
dispersions $c_r$ at which the disks reach their final,
quasi-stationary state (as a rule, after 5--10 rotations of the
outer edge of the disk) with the velocity dispersion $c_T =
3.36\pi G\sigma / \varkappa$ required to suppress gravitational
instability of a thin disk against axially symmetrical
perturbations (the Toomre criterion), as well as with other local
stability criteria. We separately analyzed the influence of the
bulge, dark halo, differential rotation, and initial disk
parameters on the radial behavior of $Q_T=c_r /c_T$.

Our numerical simulations show that differential rotation and
inhomogenity of the disk, the global nature of perturbations, and
the finite thickness of the disk can change the local velocity
dispersions (or the local Toomre parameters) of marginally stable
disks by ${\gee\! 50\%}$. The numerical models enable to carry out
a separate analysis of the effects of each of these factors.

The models considered here clearly demonstrate rapid heating of
initially unstable disks during the formation and destruction of
transient spiral arms. The low-contrast ``remnants'' of these
arms can be traced over more than ten rotations. A similar
character of disk evolution we obtained for models which used both
different numbers of particles and a different computational
algorithm (PP method).

As expected, the minimum radial-velocity dispersion at the end of
the simulations (expressed in units of the circular velocity) is
higher in the models where the relative mass of the halo is lower,
the initial disk thickness is less, and the degree of differential
disk rotation is higher. Although the radial dependence of $Q_T$
differs for different models and is determined primarily by the
relative mass and degree of concentration of the spherical
components, in all cases, $Q_T(r)$ passes through a minimum
${Q_T\simeq 1.2\div 1.6}$ at a galactocentric distance of $(1\div
2) L$, and this behavior depends only slightly on the choice of
model. This fact can be used to approximately estimate the density
(and, consequently, the mass) of a galactic disk (or to put limits
on these quantities) from observed radial-velocity dispersions
using equations~(19) and (20) without the use of numerical
simulations or analytical stability criteria. If the halo is not
too massive ($M_h/M_{\textrm{d}}\lee 2$), a disk that begins its
evolution from a strongly unstable state rather than from a
subcritical state can experience a substantial mass redistribution
over one or two rotations. The fact that galaxies usually have
exponential brightness distributions indicates that the formation
of stellar disks, apparently, was not accompanied by the
development of strong gravitational instability.

The thicker is  the disk initially, the lower is the minimum
radial velocity dispersion $c_{r}$,
 which determines its stability. Therefore, the minimum critical dispersions
 in both coordinates $z$
  and $r$ are reached if the disk begins to evolve from a subcritical
  state. It's valid
  both for gravitational perturbations in the plane and for bending perturbations
  (global, primarily axisymmetric mode, and small-scale ones).

The disk evolution traced by the numerical models clearly
demonstrates the interrelations of  processes at different $r$.
Initial stability of the central disk region slows down while
strong initial instability accelerates the increase of the
velocity dispersion at the disk periphery, having, however, no
significant effect on the final state.

Existing analytical criteria for the stability of thin disks are
based on other considerations, and, strictly speaking, there is
no reason why they must coincide with the $Q_T$ values we have
found, due to the local nature of the criteria applied, as well
as due to the use of two-dimensional analytical models, while
dynamical models treat the disk heating in three dimensions. The
model velocity dispersions differ significantly from those
predicted by various  criteria (Figs.~5b--5d, 6b, 6e, 7b--7d, 8a,
and 8b). However, it's evident that the formal application of
these criteria will yield results that are correct to order of
magnitude. In particular, like model estimates, they imply that
$Q_T > 1 $ at all $r$ and that this parameter increases at the
disk periphery. However, in none of the cases considered do the
$Q_T(r)$ relations derived using local criteria agree with those
found in this work for all radial distances ($0<r<4L$), and that
they can differ from the model estimates by more than a factor of
$1.5$ at some distances $r$.

After passing through its minimum, the Toomre parameter $Q_T(r)$
for a disk that has reached a stable equilibrium state increases
monotonically with $r$ up to $Q_T \simeq 2\div 3$ at a radius of
$(3\div 4)L$ (this radius is usually close to the optical boundary
of the disk in real galaxies). Note that a similar increase of
$Q_T(r)$ from 1.5 to 2.5 at the periphery was obtained earlier by
Pichon and Lynden-Bell~[49] using a different method (by
constructing stable equilibrium distribution functions for flat
systems over a wide range of free parameters) and has also been
mentioned for a number of dynamical models~[5, 21, 24, 44--47].

The disks of galaxies whose halo are not too massive (${\mu \lee
1.5}$) are subjected to bar instability. The suppression of the
development of a long-lived bar in galaxies with low-mass
spherical components requires not only a higher velocity
dispersion (as shown long ago by Ostriker and Peebles~[29] in
their classic work) but also a higher value of the Toomre
parameter $Q_T$, compared to galaxies possessing a massive halo or
bulge. Moreover, the development of a bar is accompanied by an
increase in $Q_T$ even in regions of the galaxy beyond the bar.
This effect is apparently associated with the system being
``overheated'' above the threshold level: the development of the
bar is accompanied by a redistribution of mass in the disk, so
that the local decrease of the disk surface density for the given
velocity dispersion results in an additional increase in $Q_T$.
However, in this case, our models fail to yield \emph{bona fide}
quantitative estimates, since the parameters of the bar depend
strongly on the initial conditions in the system. We have not
analyzed here the conditions for the development of the bar mode
in detail.

Our results make it possible to use numerical models to estimate the
degree of dynamical heating of stellar disks and to derive estimates for
the \emph{a priori} unknown mass and density of a galactic disk from the
observed velocity dispersion of the old stars making it up.

\section{ACKNOWLEDGMENTS}

We are grateful to A.M.~Fridman, O.V.~Khoruzhi{\u\i}, and V.L.~Polyachenko
for useful discussions of this work and comments. This work was supported
by the Russian Foundation for Basic Research (project no.~01-02-17597)
and partially supported by the Federal Research and Technology Program
``Research and Development in Priority Fields of Science and
Technology'' (contract 40.022.1.1.1101).

  \b=0.5\b

\hfill \emph{Translated by A. Dambis}
\newpage
\begin{figure}
\epsfbox{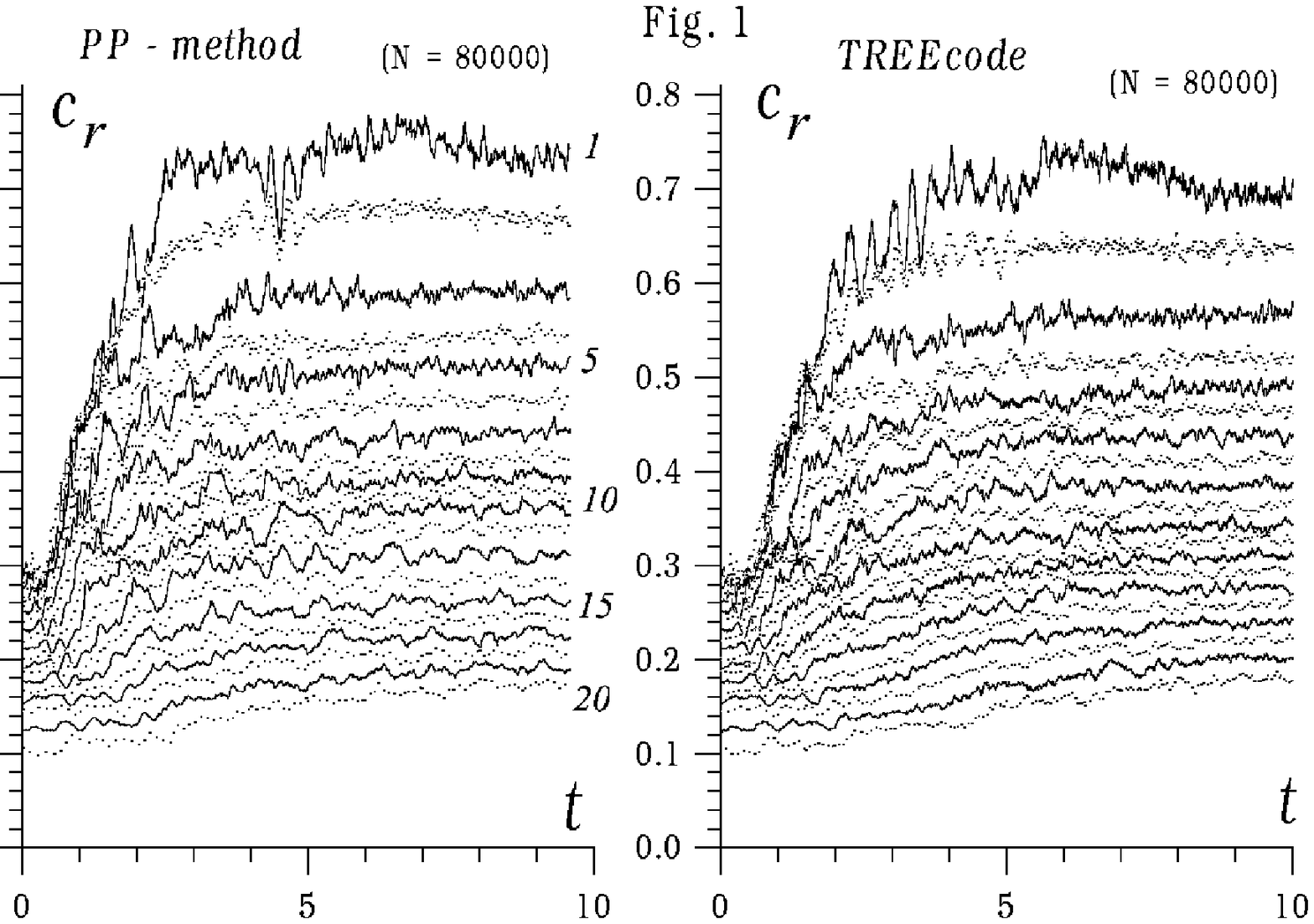}
\end{figure}

\clearpage \newpage \it{ Figure1.Time dependence of the
radial-velocity dispersion $c_r(r)$ for particles at 20 different
galactocentric distances ($1$ is the central zone, and each zone
has a width of $0.05R$) at the initial stage of heating of a
cool, thin disk ($t=10$ corresponds to $\sim\! 3$ rotations of
the outer edge of the disk) for two methods of computing the
gravitational force with ${N=80\,000}$: (a) the PP algorithm and
(b) the TREE-code algorithm.}

\newpage   \begin{figure}
\epsfbox{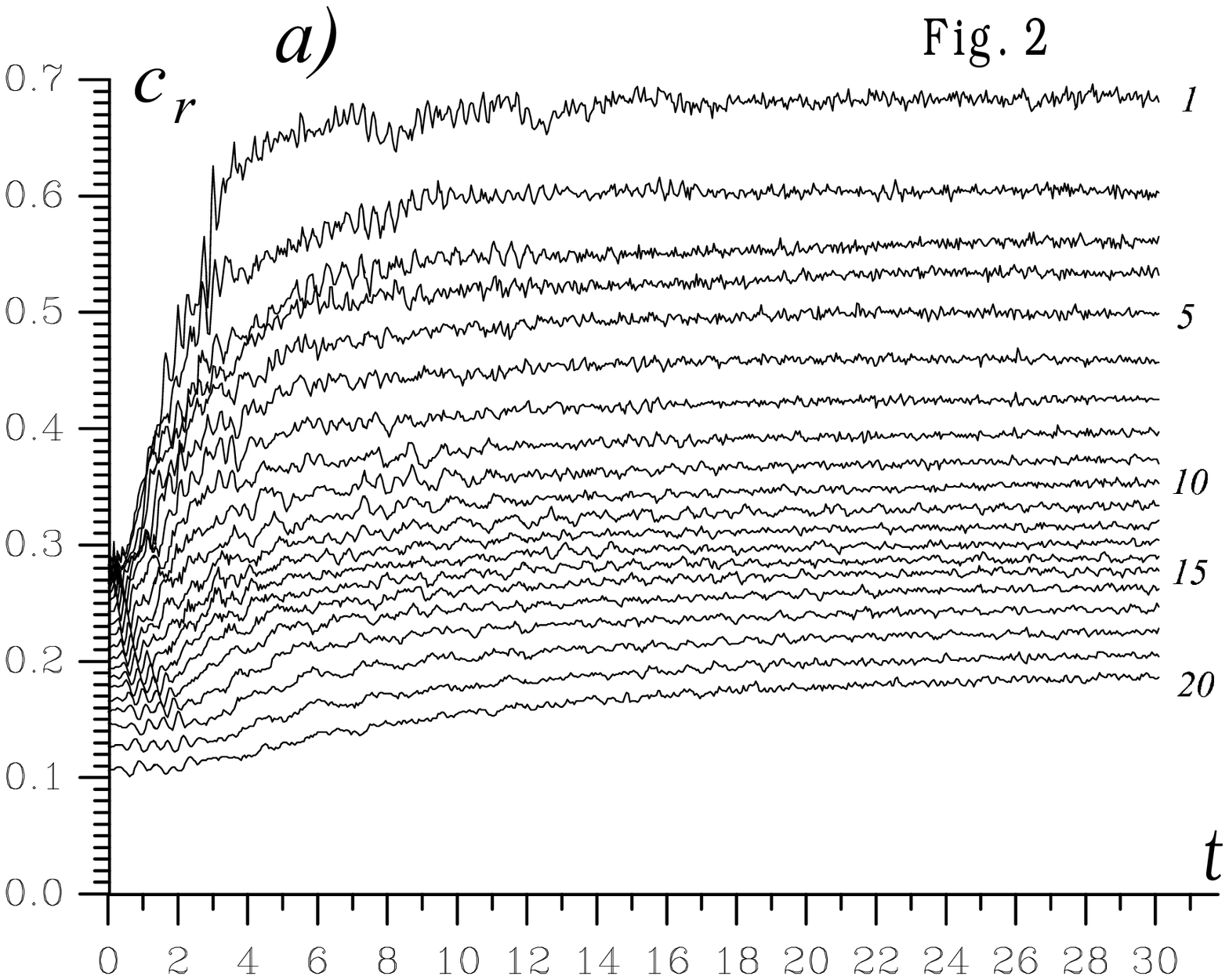}
\end{figure}
\clearpage
\newpage
\it { Figure2. Evolution of an initially unstable thin disk
($\mu=3$, $a=L$). (a) Time dependence of the radial-velocity
dispersion in 20 radial zones. (b) Development of a system of
transient spiral waves in an initially axisymmetric disk. The
positions of point masses in the ($x,y$) plane at various times
are presented (since the model consists of ${5\times 10^{5}}$
particles, only some of the point masses are shown). (c) Time
dependence of the integrated amplitudes $\hat{A}(m)$ of the
Fourier harmonics for various modes $m=2, ... , 6$. The
parameters $\hat{A}$ reach their maximum values at $1\lee t
\lee4$, and it is in this time interval that the perturbations
reach their maximum amplitudes (Fig.~1b). After $t>4$, $\hat{A}$
decreases, which corresponds to the beginning of wave dissipation
and the slowing down of the growth of $c_r$ (Fig.~2a).}

\clearpage
\newpage   \begin{figure}
\epsfbox{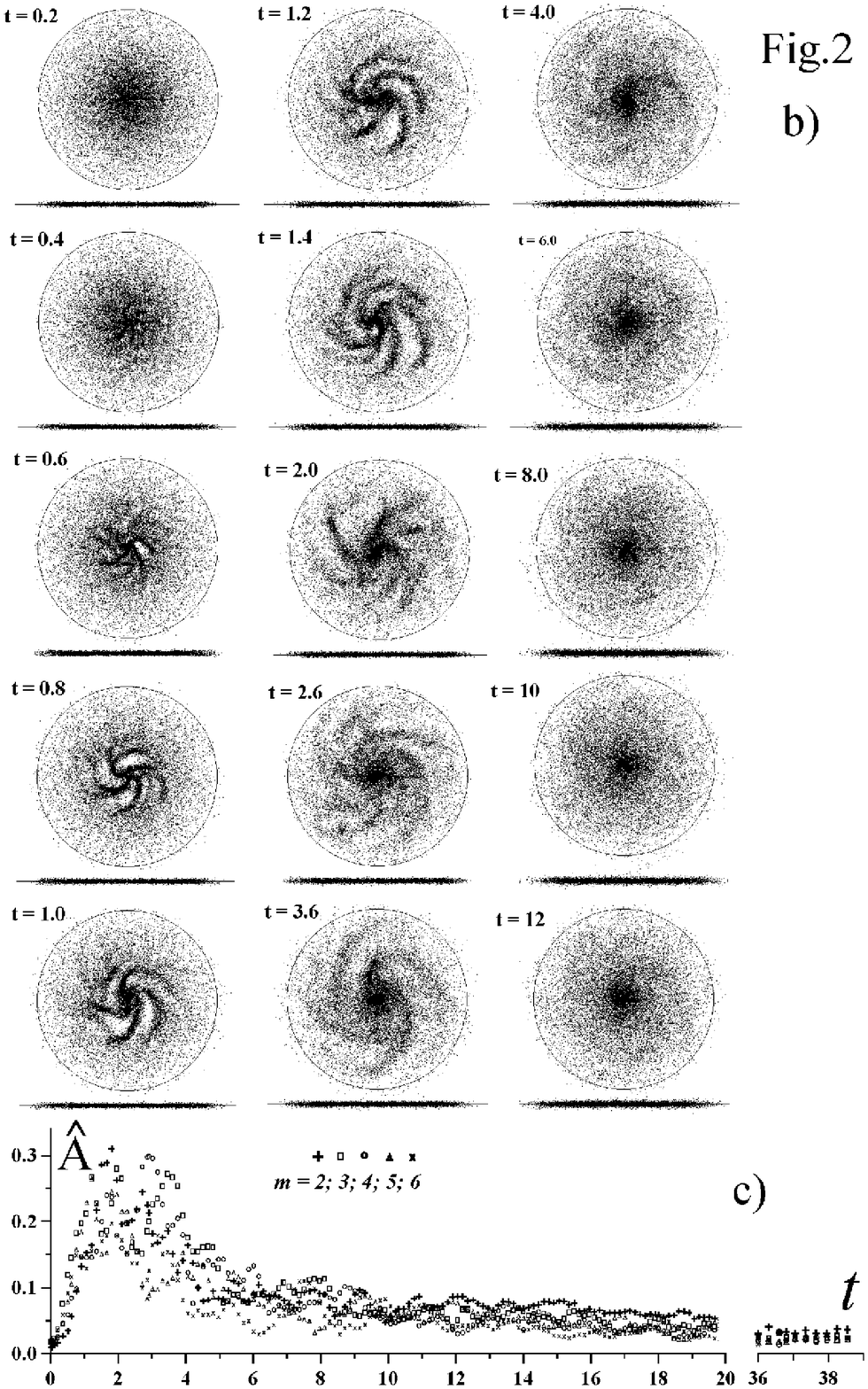}
\end{figure}

\clearpage
\newpage
\begin{figure}
\epsfbox{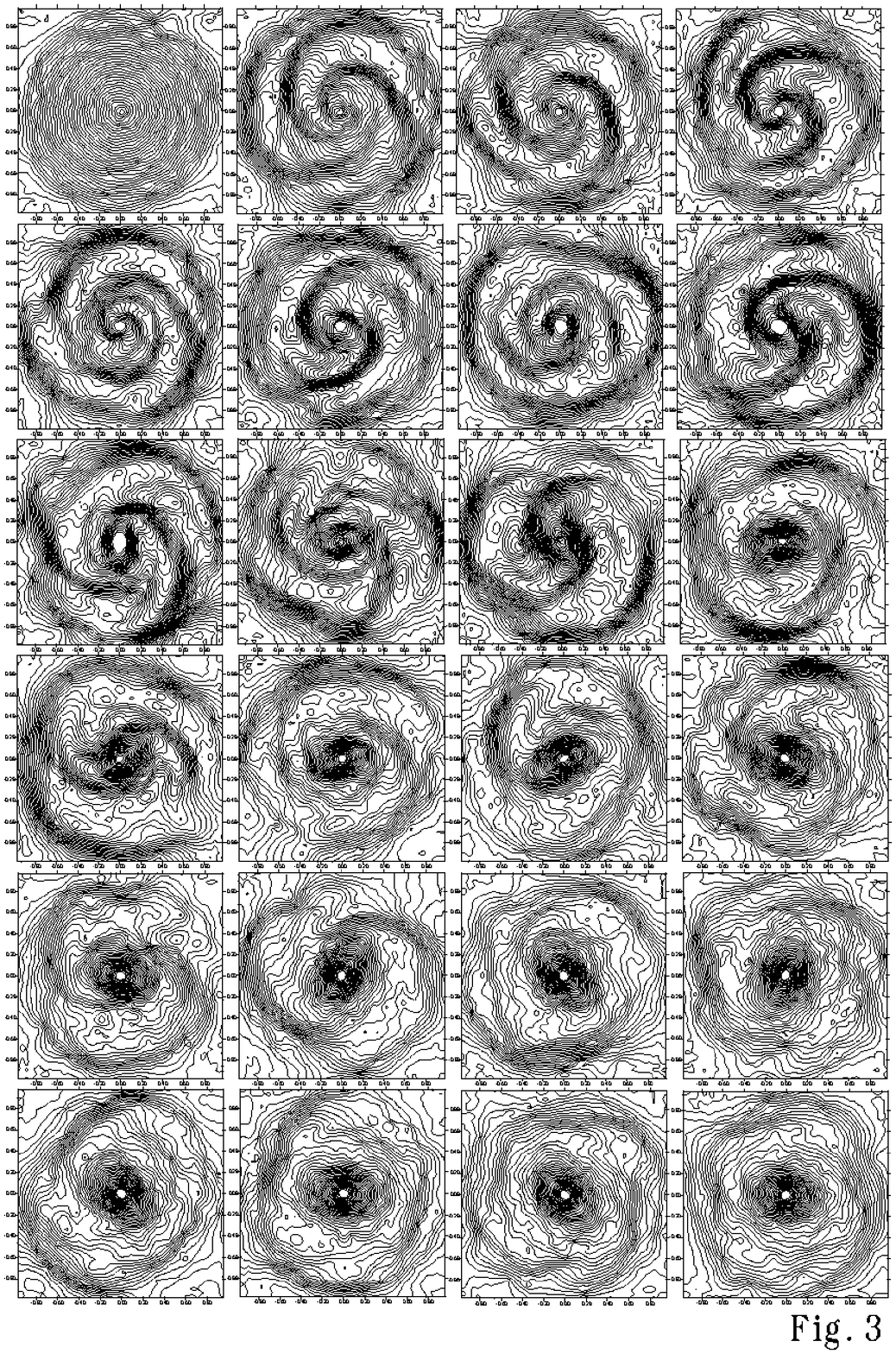}
\end{figure}
\clearpage
\newpage \it{ Figure3. Distribution of the logarithmic
surface density $\log\sigma$ at various times from $t=1$ to
$t=30$ for the initial subcritical state with ${Q_T\gee1}$.}

\newpage   \begin{figure}
\epsfbox{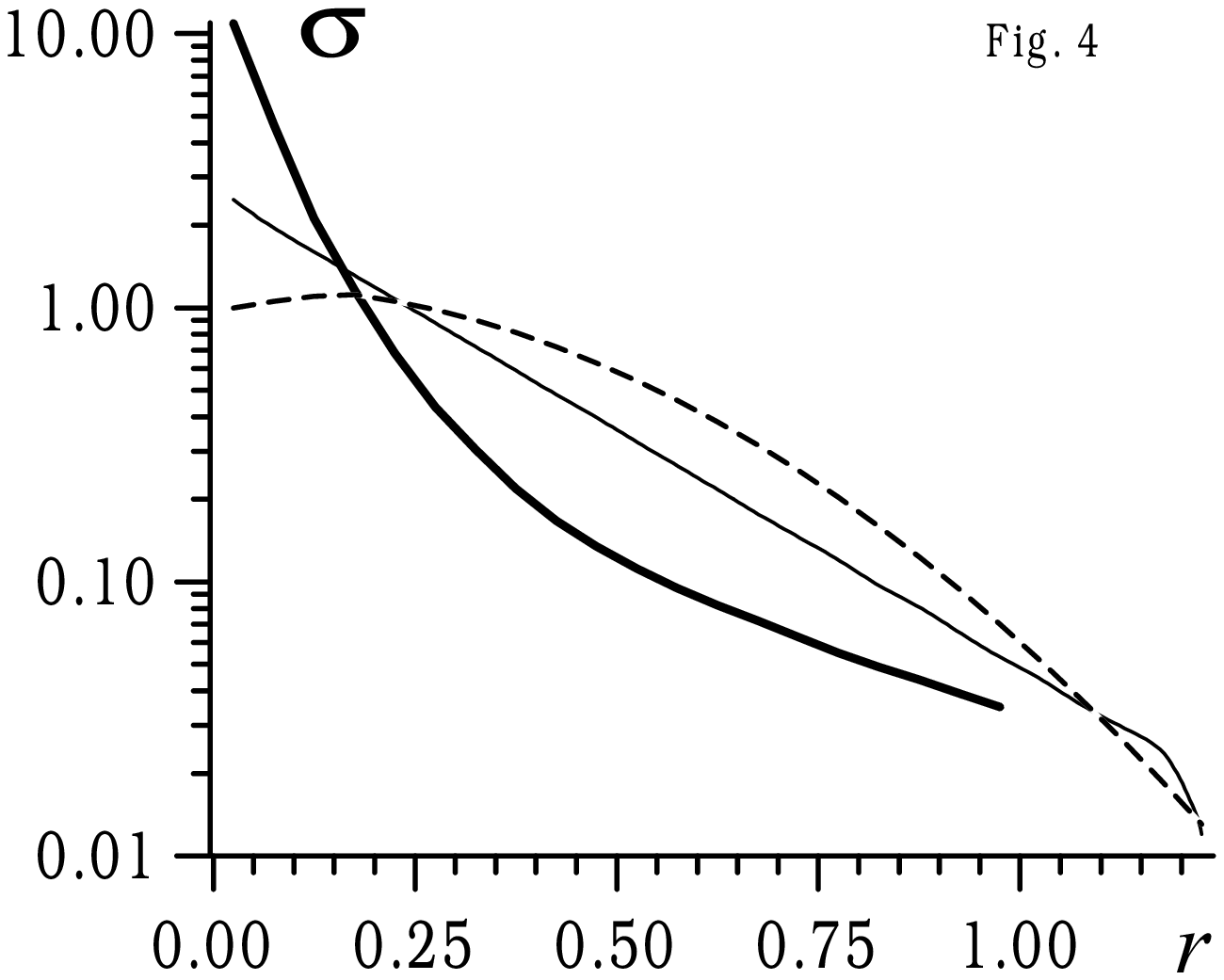}
\end{figure}
\clearpage \newpage \it {
Fig4. Radial dependences of the surface
density for an initially cool model disk in which there is
subsequently a substantial density redistribution. The thin solid
curve shows an exponential profile, the bold solid curve is the
final profile averaged in azimuth after strong heating and
development of a bar, and the dashed curve is the initial profile
that produces a quasi-exponential distribution at the end of the
computations. \hfill}

\newpage   \begin{figure}
\epsfbox{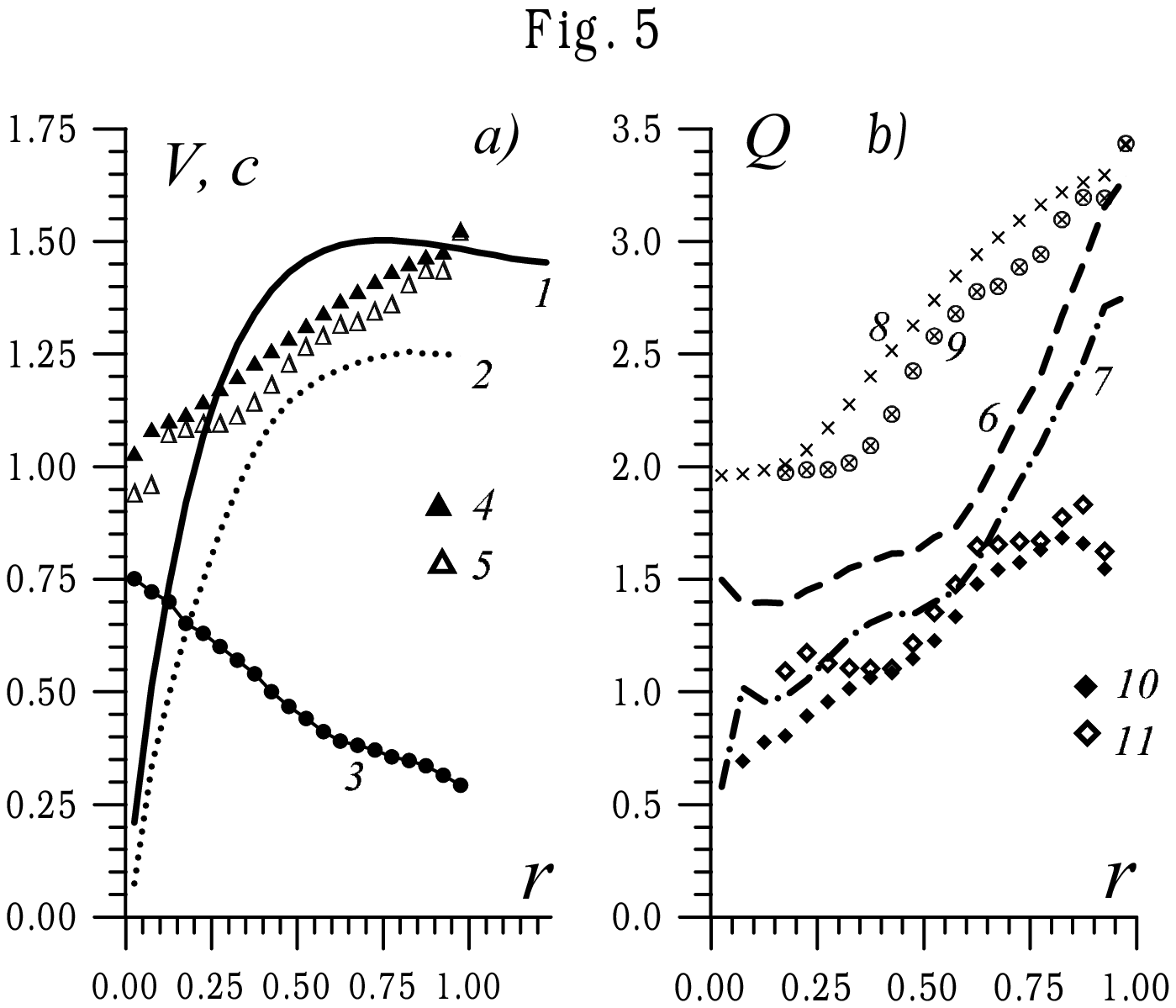}
\end{figure}

\clearpage \newpage
\it {
Fig.5. Parameters of the disk that has evolved
to a stable state at the stability limit, for bulgeless models.
(a) Radial dependences of the circular velocity $V_c(r)$ (curve
\emph{1}), disk rotational velocity $V(r)$ (curve~\emph{2}),
radial-velocity dispersion $c_r(r)$ (curve~\emph{3}), and
parameters $S$ = $2\Omega /\varkappa$ computed for the circular
velocity (curve~\emph{4}) and disk rotational velocity
(curve~\emph{5}), respectively. Results are shown for the model
in which the mass of the halo within $r\leq 1 =4L$ is equal to
that of the disk and the halo scale length $a$ is equal to the
radial disk scale length $L$. (b) Radial dependences of the
Toomre parameter at the stability limit computed using various
stability criteria for the model shown in Fig.~5a: \emph{6} $Q_T$
determined from $V_c$, \emph{7} $Q_T$ determined from $V$,
\emph{8} $Q_T^{(P)}$ determined from $V_c$, \emph{9} $Q_T^{(P)}$
determined from $V$, \emph{10} $Q_T^{(M)}$ determined from $V_c$,
and \emph{11} $Q_T^{(M)}$ determined from $V$.}

\newpage   \begin{figure}
\epsfbox{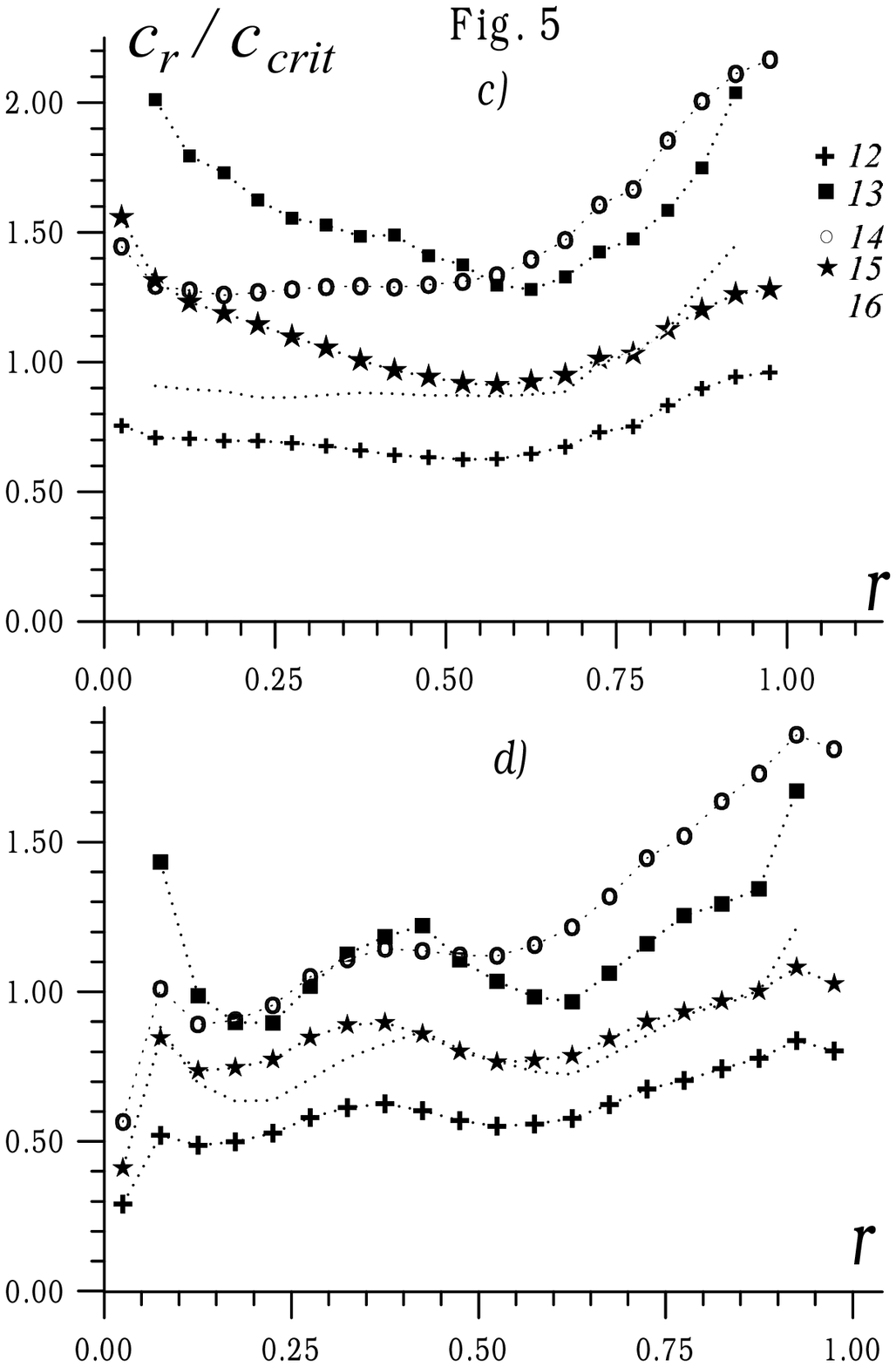}
\end{figure}
\clearpage
\newpage
\it {
Fig.5(cont.)(c) Radial distributions of the
ratio $c_r/c_{\textrm{crit}}$ of the radial-velocity dispersions
in the numerical models (curve \emph{3} in Fig.~5a) to the
critical values computed using the circular rotational velocity
$V_c(r)$ (curve~\emph{1} in Fig.~5a) for \emph{12} the
Polyachenko--Polyachenko--Strel'nikov criterion (6) and \emph{13}
the Morozov criterion (5). Also shown are \emph{14}
$c_r/(c_T2\Omega/\varkappa)$ derived using the simplified
criterion (4); \emph{15} $c_r$ normalized to $c^{(P)}Q_T^{(1)}$,
which is a generalization of criterion~(6) for the case of a
finite-thickness disk, derived using~(2); and \emph{16} the ratio
obtained for the Morozov criterion (5) in the case of an
infinitely thin disk (${D=1}$). (d) Same as Fig.~5c using the
disk rotational velocity $V(r)$ (see curve~\emph{2} in Fig.~5a).}

\newpage   \begin{figure}
\epsfbox{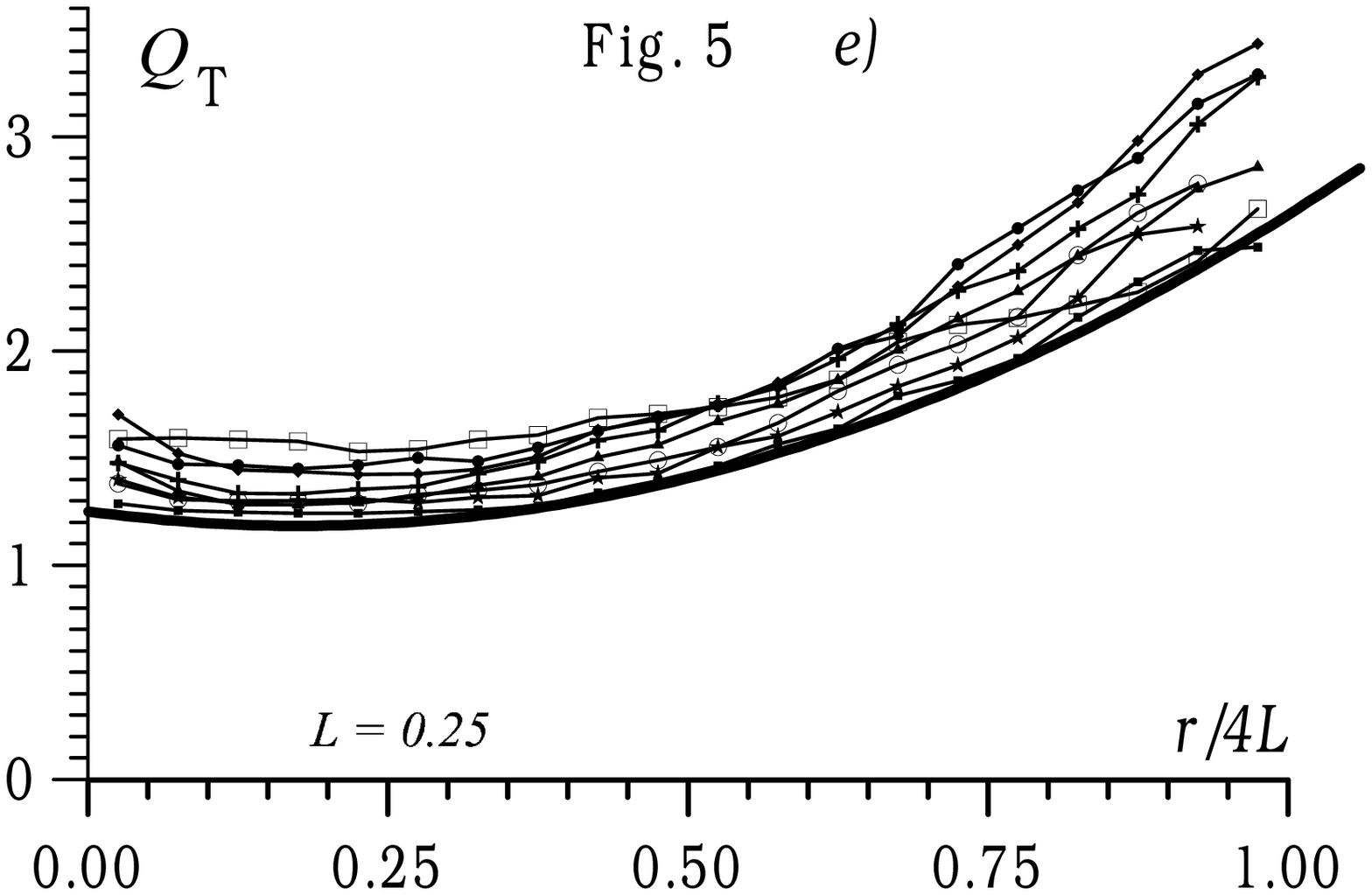}
\end{figure}

\clearpage
\newpage
\it {
 Fig.5(cont.) (e) The
Toomre parameter $Q_T(r)$ at the stability limit for a series of
bulgeless models with various halo parameters. The bold thick
curve is computed using~(19).}

\newpage   \begin{figure}
\epsfbox{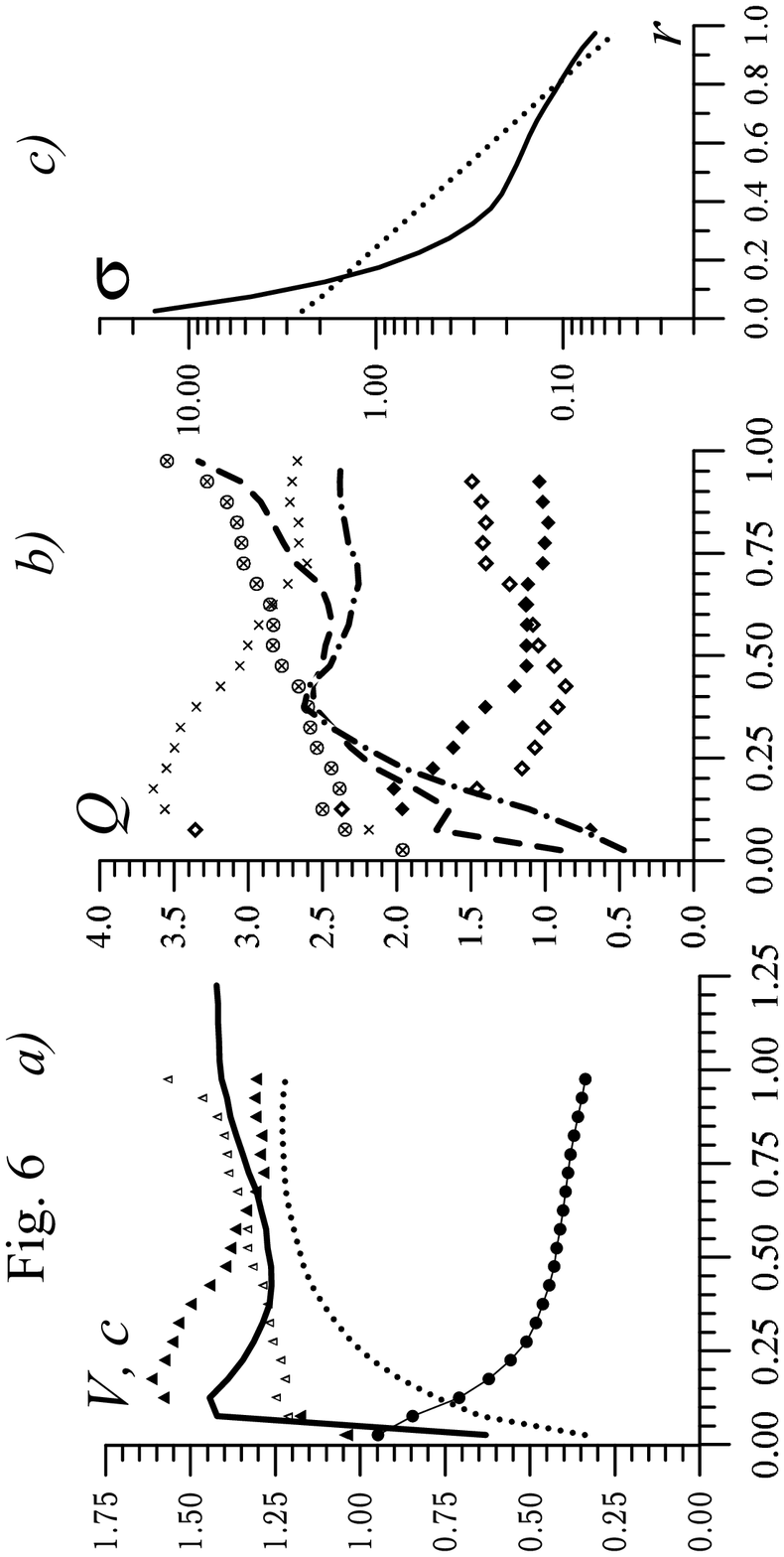}
\end{figure}

\clearpage \newpage \it { Fig.6. Bulgeless models that develop
bars in the course of their evolution. (a)--(c)
$M_h=0.7M_{\textrm{d}}$, and $a=2L$;
 (d)--(f)  $M_h=2M_{\textrm{d}}$ and $a = 1.6L$. Same
notation as in Fig.~5. Plots  c  and  f  show azimuthally
averaged surface-density profiles at the initial time (dashed)
and after the development of a quasi-stationary state with a bar
at $t\geq 25$ (solid).}

\newpage
 \begin{figure}
\epsfbox{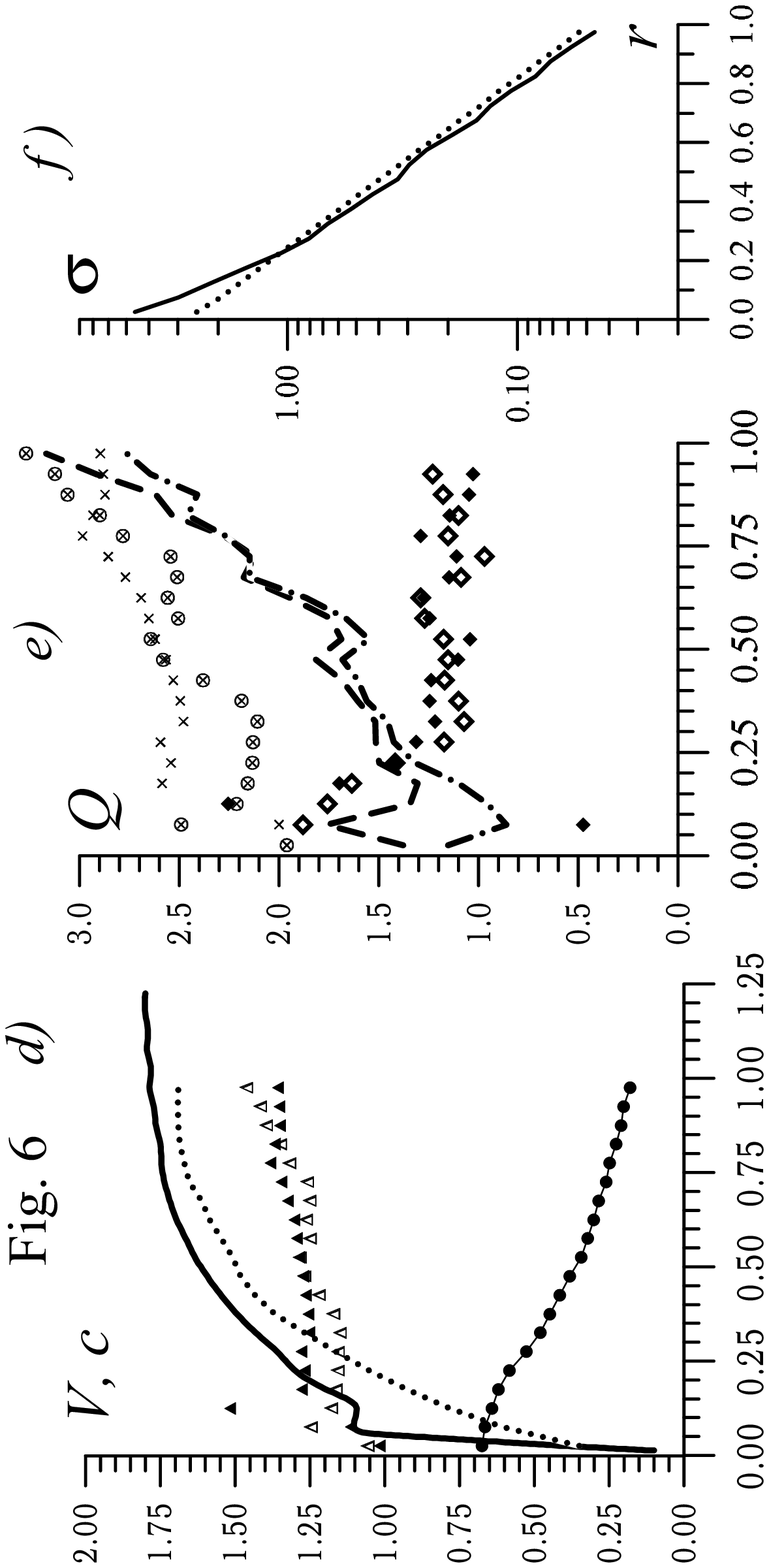}
\end{figure}

\newpage   \begin{figure}
\epsfbox{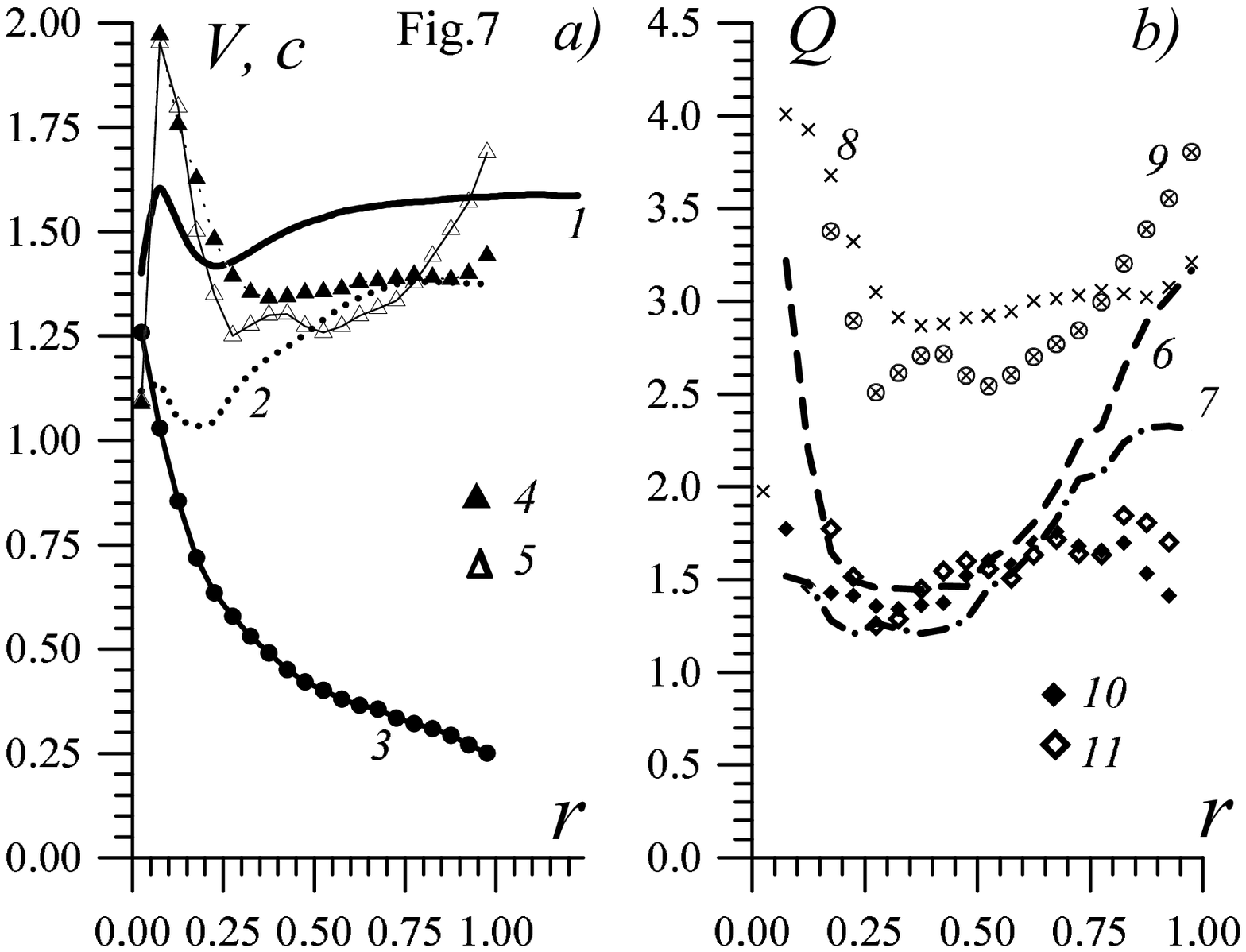}
\end{figure}

\clearpage
\newpage
\it {
Figure7. Disk parameters at the stability limit for models with
bulges. (a)--(e)  Radial dependences of the disk parameters (same
notation as in Fig.~5). The mass of the halo within $r\leq 1=4L$
is equal to that of the disk, and the halo scale length is
$a=3.6L$. The bulge parameters are $M_b=0.24M_{\textrm{d}}$,
$b=0.04L$, and $(r_b)_{\textrm{max}}= 0.5L$.  (f)  Toomre
parameter $Q_T(r)$ computed using $V_c(r)$ at the stability limit
for the series of models with bulges. The bold solid curve is
based on~(19). (g) Edge-on view of the disk at the end of the
computation of the model shown in \emph{a}. The dots show the
positions of the particles. The model develops an appreciable
bulge-like feature in the central region of the disk.\hfill}

\newpage   \begin{figure}
\epsfbox{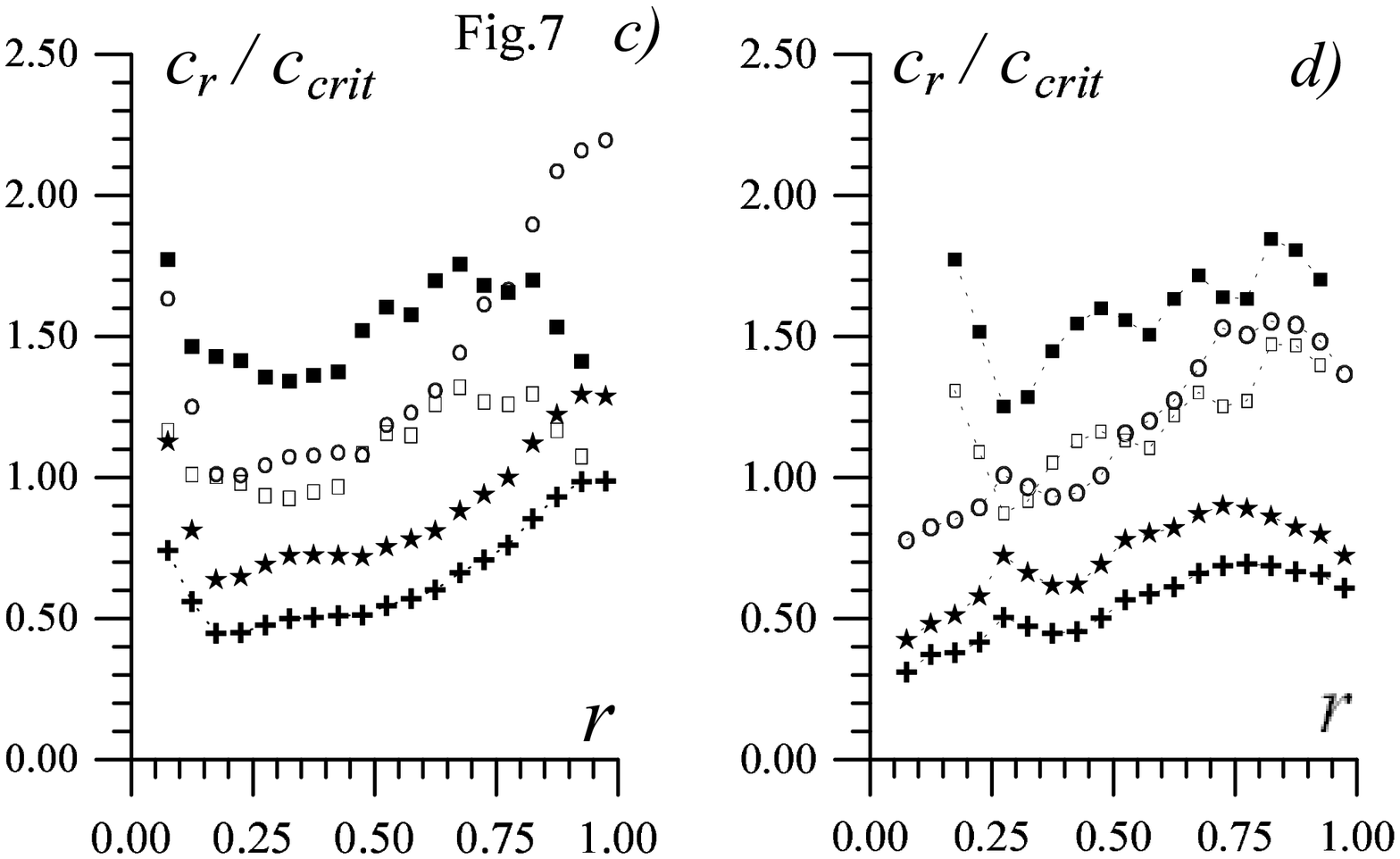} 
\end{figure}

\newpage   \begin{figure}
\epsfbox{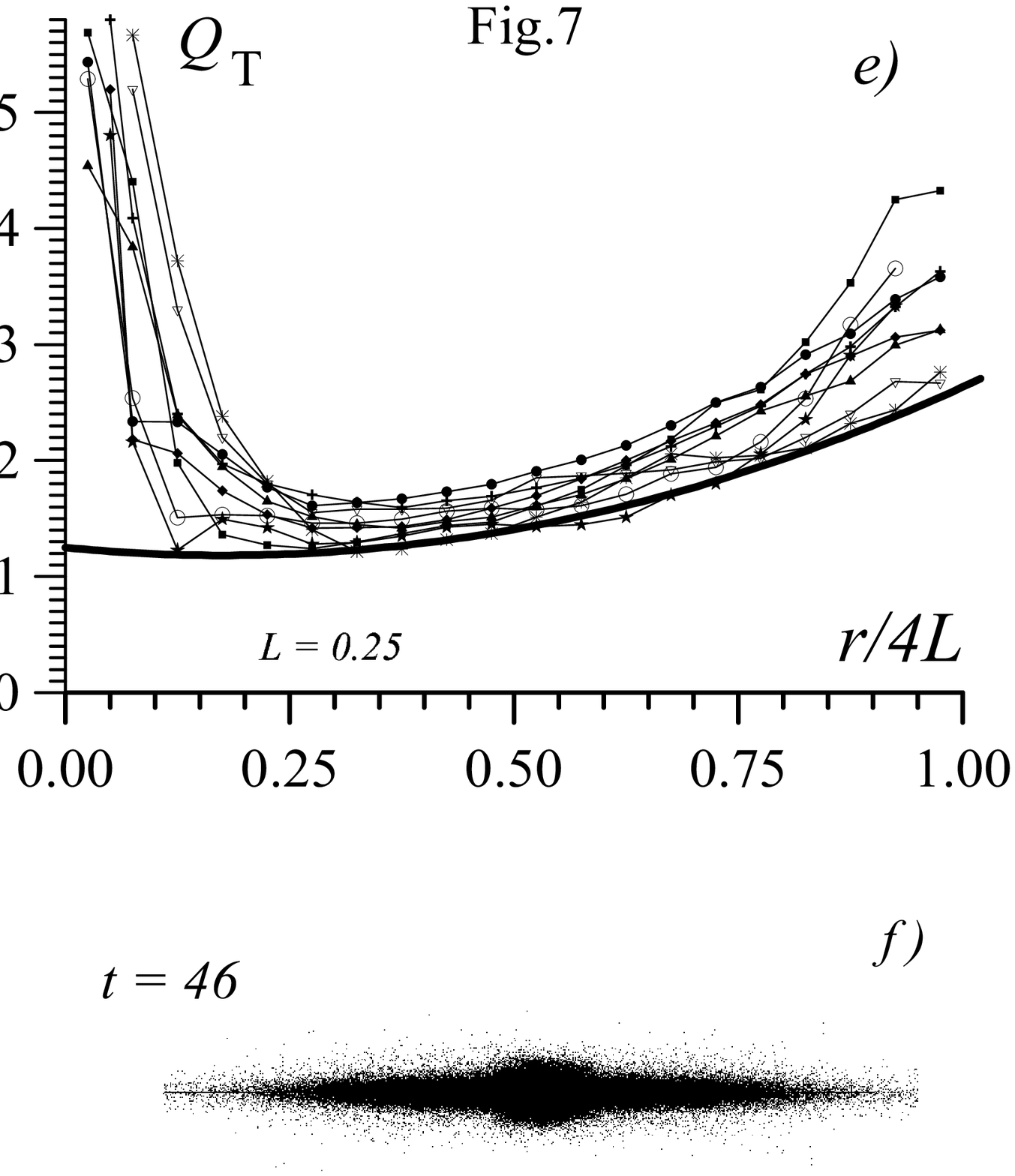} 
\end{figure}

\newpage   \begin{figure}
\epsfbox{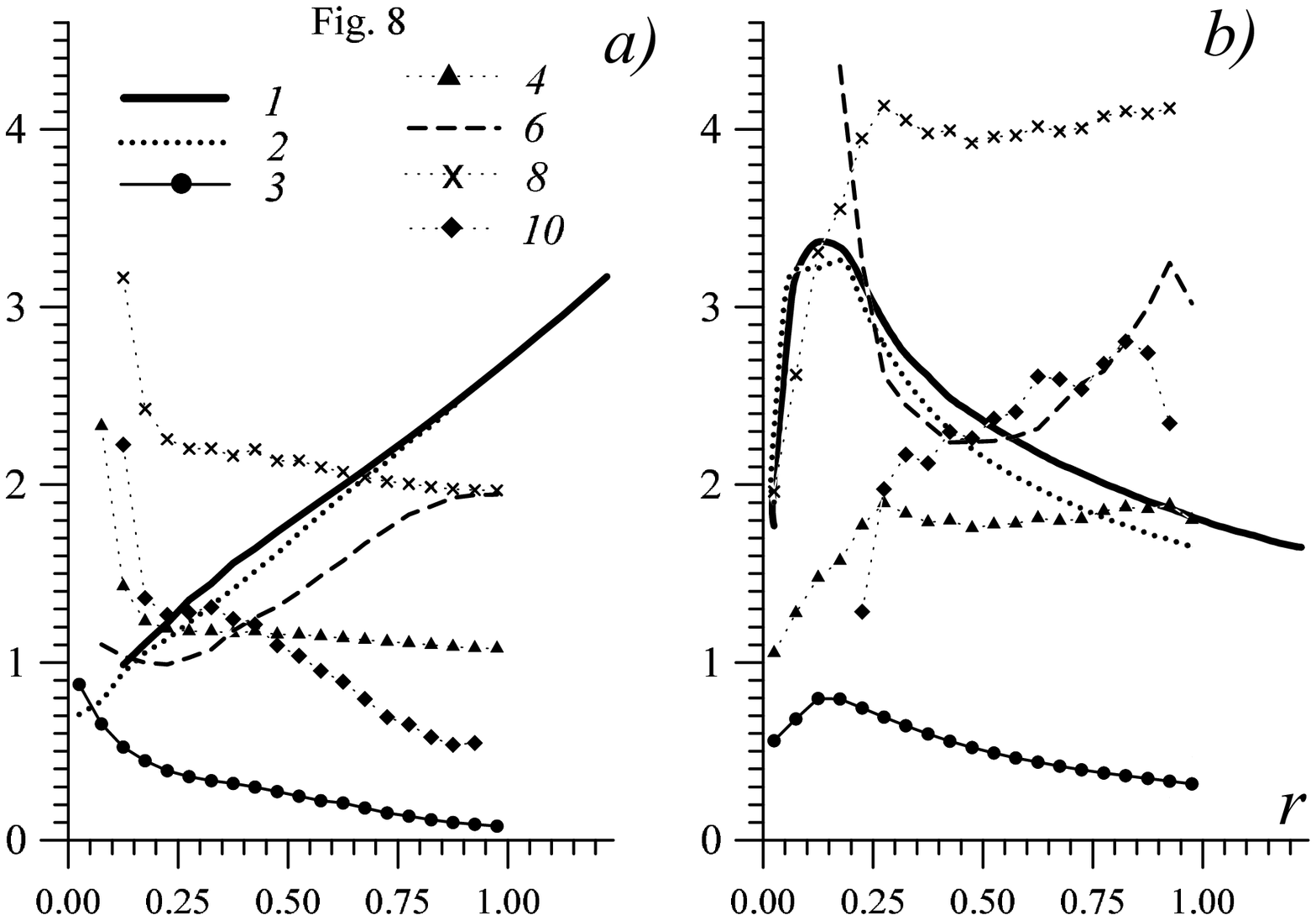}
\end{figure}
\clearpage
\newpage
\it {
Fig.8 Radial dependences of the disk
parameters in the adopted units for the cases when (a) a
significant part of the disk rotates in accordance with a
quasi-rigid law, and (b) the rotation at $r>L=0.25$ is close to
Keplerian: \emph{1} the circular velocity $V_c(r)$; \emph{2} the
rotational velocity of the disk (stars) $V(r)$; \emph{3} the
radial-velocity dispersion $c_r(r)$; \emph{4} the parameter $S$
computed using~(3) for the circular velocity; \emph{6} the Toomre
parameter $Q_T$ computed using the circular velocity; \emph{8}
the critical Toomre parameter $Q_T^{(P)}$ computed using~(6) with
$V_c(r)$; \emph{10} the critical Toomre parameter $Q_T^{(M)}$
computed using~(5) with $V_c(r)$. The notation is analogous to
that in Fig.~5. \hfill}

\begin{thebibliography}{99}

\bibitem{1:Khoperskov_n}
A.~Toomre, Astrophys. J. \textbf{139}, 1217 (1964).

\bibitem{2:Khoperskov_n}
A.~G.~Morozov, Astron. Zh. \textbf{58}, 734 (1981) [Sov. Astron.
Rep. \textbf{25}, 421 (1981)].

\bibitem{3:Khoperskov_n}
A.~V.~Zasov, Pis'ma Astron. Zh. \textbf{11}, 730 (1985) [Sov.
Astron. Lett. \textbf{11}, 307 (1985)].

\bibitem{4:Khoperskov_n}
R.~Bottema, Astron. Astrophys. \textbf{275}, 16 (1993).

\bibitem{5:Khoperskov_n}
R.~Bottema and J.~P.~E.~Gerritsen, Mon. Not. R. Astron. Soc.
\textbf{290}, 585 (1997).

\bibitem{6:Khoperskov_n}
B.~Fuchs, in \emph{Galaxy Dynamics}, ASP Conf. Ser. \textbf{182}, Ed.
by R.~D.~Merritt, M.~Valluri, and J.~A.~Sellwood (Astronomical Society
of the Pacific, San Francisco, 1999), p. 365.

\bibitem{7:Khoperskov_n}
A.~V.~Khoperskov, A.~V.~Zasov, and N.~V.~Tyurina, Astron. Zh.
\textbf{78}, 213 (2001) [Astron. Rep. \textbf{45}, 180 (2001)];
A.~V.~Zasov, D.~V.~Bizyaev, D.~I.~Makarov, and N.~V.~Tyurina,
Pis'ma Astron. Zh. \textbf{28}, 599 (2002) [Astron. Lett.
\textbf{28}, 527 (2002)].

\bibitem{8:Khoperskov_n}
A.~M.~Fridman and V.~L.~Polyachenko, \emph{Physics of Gravitating
Systems} (Springer-Verlag, New York, 1984).

\bibitem{9:Khoperskov_n}
V.~L.~Polyachenko and A.~M.~Fridman, \emph{Equilibrium and
Stability of Gravitating Systems} [in Russian] (Nauka, Moscow, 1976).

\bibitem{10:Khoperskov_n}
N.~N.~Gor'kavy{\u\i} and A.~M.~Fridman, \emph{Physics of Planetary
Rings} [in Russian] (Nauka, Moscow, 1994).

\bibitem{11:Khoperskov_n}
C.~C.~Lin and F.~H.~Shu, Astrophys. J. \textbf{140}, 646 (1964).

\bibitem{12:Khoperskov_n}
O.~P.~Vandervoort, Astrophys. J. \textbf{161}, 87 (1970).

\bibitem{13:Khoperskov_n}
A.~G.~Morozov, Astron. Zh. \textbf{57}, 681 (1980) [Sov. Astron.
Rep. \textbf{24}, 397 (1980)].

\bibitem{14:Khoperskov_n}
A.~G.~Morozov and A.~V.~Khoperskov, Astrofizika \textbf{24}, 467
(1986).

\bibitem{15:Khoperskov_n}
E.~Griv, C.~Yuan, and M.~Gedalin, Mon. Not. R. Astron. Soc.
\textbf{307}, 1 (1999).

\bibitem{16:Khoperskov_n}
A.~G.~Morozov, Pis'ma Astron. Zh. \textbf{7}, 197 (1981) [Sov.
Astron. Lett. \textbf{7}, 109 (1981)].

\bibitem{17:Khoperskov_n}
V.~L.~Polyachenko, E.~V.~Polyachenko, and A.~V.~Strel'nikov,
Pis'ma Astron. Zh. \textbf{23}, 598 (1997) [Astron. Lett.
\textbf{23}, 525 (1997)].

\bibitem{18:Khoperskov_n}
N.~W.~Evans and J.~C.A.~Read, Mon. Not. R. Astron. Soc.
\textbf{300}, 83 (1998а).

\bibitem{19:Khoperskov_n}
N.~W.~Evans and J.~C.~A.~Read, Mon. Not. R. Astron. Soc.
\textbf{300}, 106 (1998б).

\bibitem{20:Khoperskov_n}
J.~A.~Sellwood and N.~W.~Evans, Astrophys. J. \textbf{546}, 176
(2001).

\bibitem{21:Khoperskov_n}
G.~Bertin, C.~C.~Lin, S.~A.~Lowe, and R.~P.~Thurstans, Astrophys.
J. \textbf{338}, 78 (1989).

\bibitem{22:Khoperskov_n}
R.~H.~Miller, Astrophys. J. \textbf{223}, 811 (1978).

\bibitem{23:Khoperskov_n}%
R.~H.~Miller, K.~H.~Prendergast, and W.~J.~Quirk, Astrophys. J.
\textbf{161}, 903 (1970).

\bibitem{24:Khoperskov_n}
F.~Hohl, Astrophys. J. \textbf{168}, 343 (1971).

\bibitem{25:Khoperskov_n}
F.~Hohl, Astron. Space Sci. \textbf{14}, 91 (1971).

\bibitem{26:Khoperskov_n}
R.~H.~Miller, Astron. Space Sci. \textbf{14}, 73 (1971).

\bibitem{27:Khoperskov_n}
R.~H.~Miller, Astrophys. J. \textbf{190}, 539 (1974).

\bibitem{28:Khoperskov_n}
R.~H.~Miller, Astrophys. J. \textbf{223}, 811 (1978).

\bibitem{29:Khoperskov_n}
J.~P.~Ostriker and P.~J.E.~Peebles, Astrophys. J. \textbf{186},
467 (1973).

\bibitem{30:Khoperskov_n}
R.~A.~James and J.~A.~Sellwood, Mon. Not. R. Astron. Soc.
\textbf{182}, 331 (1978).

\bibitem{31:Khoperskov_n}
V.~L.~Polyachenko and I.~G.~Shukhman, Preprint Nos. 1--2,
SIBIZMIR, Sib. Div. Akad. Nauk SSSR (1972).

\bibitem{32:Khoperskov_n}
A.~J.~Kalnajs, Astrophys. J. \textbf{175}, 63 (1972).

\bibitem{33:Khoperskov_n}
E.~A.~Mikha\u{i}lova and A.~V.~Khoperskov, Astron. Zh. \textbf{69},
1112 (1992) [Astron. Rep. \textbf{36}, 573 (1992)].

\bibitem{34:Khoperskov_n}
E.~Athanassoula and J.~A.~Sellwood, Mon. Not. R. Astron. Soc.
\textbf{221}, 213 (1986).

\bibitem{35:Khoperskov_n}
C.~J.~Jog and P.~M.~Solomon, Astrophys. J. \textbf{276}, 114
(1984).

\bibitem{36:Khoperskov_n}
C.~J.~Jog and P.~M.~Solomon, Astrophys. J. \textbf{276}, 127
(1984).

\bibitem{37:Khoperskov_n}
B.~Wang and J.~Silk, Astrophys. J. \textbf{427}, 759 (1994).

\bibitem{38:Khoperskov_n}
V.~G.~Ortega, E.~Volkov, and L.~Monte-Lima, Astron. Astrophys.
\textbf{366}, 276 (2001).

\bibitem{39:Khoperskov_n}
J.~A.~Sellwood and R.~G.~Carlberg, Astrophys. J. \textbf{282}, 61
(1984).

\bibitem{40:Khoperskov_n}
R.~G.~Carlberg and J.~A.~Sellwood, Astrophys. J. \textbf{292}, 79
(1985).

\bibitem{41:Khoperskov_n}
J.~A.~Sellwood and E.~Athanassoula, Mon. Not. R. Astron. Soc.
\textbf{221}, 195 (1986).

\bibitem{42:Khoperskov_n}
J.~A.~Sellwood, Mon. Not. R. Astron. Soc. \textbf{238}, 115
(1989).

\bibitem{43:Khoperskov_n}
J.~A.~Sellwood and D.~N.C.~Lin, Mon. Not. R. Astron. Soc.
\textbf{240}, 991 (1989).

\bibitem{44:Khoperskov_n}
J.~A.~Sellwood and E.~Athanassoula, in \emph{Internal Kinematics
and Dynamics of Galaxies,} (Reidel, Dordrecht, 1983), p.~203.

\bibitem{45:Khoperskov_n}
H.~C.~Schroeder and N.~F.~Comins, Astrophys. J. \textbf{346}, 108
(1989).

\bibitem{46:Khoperskov_n}
B.~Fuchs and S.von Linden, Mon. Not. R. Astron. Soc. \textbf{294},
513 (1998).

\bibitem{47:Khoperskov_n}
J.~M.~Bardeen, \emph{IAU Symp. 69: Dynamics of Stellar Systems,} Ed. by
A.~Hayli (D.~Reidel, Dordrecht, 1975), p. 297.

\bibitem{48:Khoperskov_n}
A.~V.~Zasov, A.~G.~Morozov, Astron. Zh. \textbf{62}, 475 (1985)
[Sov. Astron. Rep. \textbf{29}, 277 (1985)].

\bibitem{49:Khoperskov_n}
C.~Pichon and D.~Lynden-Bell, Mon. Not. R. Astron. Soc.
\textbf{282}, 1143 (1996).

\bibitem{50:Khoperskov_n}
A.~B.~Romeo, Astron. Astrophys. \textbf{286}, 799 (1994).

\bibitem{51:Khoperskov_n}
A.~B.~Romeo, Astron. Astrophys. \textbf{335}, 922 (1998).

\bibitem{52:Khoperskov_n}
J.~Sommer-Larsen, H.~Vedel, and U.~Hellsten, Mon. Not. R. Astron.
Soc. \textbf{294}, 485 (1998).

\bibitem{53:Khoperskov_n}
J.~C.~Lambert and A.~Bosma, Mon. Not. R. Astron. Soc.
\textbf{314}, 475 (2000).

\bibitem{54:Khoperskov_n}
J.~N.~Bahcall, Astrophys. J. \textbf{276}, 156 (1984).

\bibitem{55:Khoperskov_n}
E.~Athanassoula and A.~Misiriotis, Mon. Not. R. Astron. Soc.
\textbf{330}, 35 (2002).

\bibitem{56:Khoperskov_n}
V.~P.~Debattista and J.~A.~Sellwood, Astrophys. J. \textbf{543},
704 (2000).

\bibitem{57:Khoperskov_n}
H.~Hasan and C.~Norman, Astrophys. J. \textbf{361}, 69 (1990).
\end{thebibliography}
\end{document}